\documentclass[preprint,11pt]{JHEP3} 

\JHEPspecialurl{http://jhep.sissa.it/JOURNAL/JHEP3.tar.gz}

\usepackage{epsfig,multicol,amsmath}

\def\beq{\begin{equation}}
\def\eeq{\end{equation}}
\def\bea{\begin{eqnarray}}
\def\eea{\end{eqnarray}}

\def\eq#1{{Eq.~(\ref{#1})}}

\def\fig#1{{Fig.~\ref{#1}}}
\def\sec#1{{section~\ref{#1}}}
\def\rho{x}
\newcommand{\bas}{\bar{\alpha}_s}
\newcommand{\as}{\alpha_s}

\newcommand{\mn}{{\mu\nu}}

\newcommand{\Lb}{\left(}
\newcommand{\Rb}{\right)}

\parskip=\itemsep               

\newcommand{\p}{I\!\!P}

\renewcommand{\theequation}{\thesection.\arabic{equation}}

\def\ap#1#2#3{     {\it Ann. Phys. (NY) }{\bf #1} (19#2) #3}
\def\arnps#1#2#3{  {\it Ann. Rev. Nucl. Part. Sci. }{\bf #1} (19#2) #3}
\def\npb#1#2#3{    {\it Nucl. Phys. }{\bf B#1} (19#2) #3}
\def\plb#1#2#3{    {\it Phys. Lett. }{\bf B#1} (19#2) #3}
\def\prd#1#2#3{    {\it Phys. Rev. }{\bf D#1} (19#2) #3}
\def\prep#1#2#3{   {\it Phys. Rep. }{\bf #1} (19#2) #3}
\def\prl#1#2#3{    {\it Phys. Rev. Lett. }{\bf #1} (19#2) #3}
\def\ptp#1#2#3{    {\it Prog. Theor. Phys. }{\bf #1} (19#2) #3}
\def\rmp#1#2#3{    {\it Rev. Mod. Phys. }{\bf #1} (19#2) #3}
\def\zpc#1#2#3{    {\it Z. Phys. }{\bf C#1} (19#2) #3}
\def\mpla#1#2#3{   {\it Mod. Phys. Lett. }{\bf A#1} (19#2) #3}
\def\nc#1#2#3{     {\it Nuovo Cim. }{\bf #1} (19#2) #3}
\def\yf#1#2#3{     {\it Yad. Fiz. }{\bf #1} (19#2) #3}
\def\cpc#1#2#3{    {\it Comp. Phys. Commun. }{\bf #1} (19#2) #3}
\def\dis#1#2{      {\it Dissertation, }{\sf #1 } 19#2}
\def\dip#1#2#3{    {\it Diplomarbeit, }{\sf #1 #2} 19#3 }
\def\ib#1#2#3{     {\it ibid. }{\bf #1} (19#2) #3}
\def\jpg#1#2#3{        {\it J. Phys}. {\bf G#1}#2#3}
\title{\LARGE \bf Electromagnetic Higgs production}
\author{\large  J.~Miller\thanks{Email:
jeremymi@post.tau.ac.il;}\,\, \\
Department of Particle Physics, School of Physics and Astronomy\\
Raymond and Beverley Sackler
 Faculty
of Exact Science\\  Tel Aviv University, Tel Aviv, 69978, Israel}

\abstract{The cross section for central diffractive Higgs production
is calculated, for the LHC range of energies. The graphs for the
possible mechanisms for Higgs production, through pomeron fusion and
photon fusions are calculated for all possibilities allowed by the
standard model. The cross section for central diffractive Higgs
production through pomeron fusion, must be multiplied by a factor
for the survival probability, to isolate the Higgs signal and reduce
the background. Due to the small value of the survival probability
$\Lb\,4\,\times\, 10^{-3}\Rb\,$, the cross sections for central
diffractive Higgs production, in the two cases for pomeron fusion
and photon fusion, are competitive.}


\preprint{ \today}
\begin{document}

\section{Introduction}
\label{sec:Int}

The most promising process for observation of the Higgs boson at the
LHC  is central diffractive production of the Higgs, with large
rapidity gaps ( LRG ) between the Higgs and the two emerging
protons, after scattering. Namely,

\beq
p\,+\,p\,\rightarrow\,p\,+\,[\mbox{LRG}\,]\,+\,H\,+\,[\,\mbox{LRG}\,]\,+p\label{ppH}
\eeq

The large rapidity gaps either side of the Higgs reduces the
background, so that the Higgs signal will be easier to isolate in
central exclusive production. Hence, this process gives the best
experimental signature for detecting the Higgs at the LHC, and is
very interesting for experiments in the search for the Higgs
boson.\\

In this paper, two mechanisms are compared for central exclusive
Higgs production; (1)
 $\gamma\,\gamma$ fusion, namely $pp\rightarrow\,\gamma\,\gamma\,\rightarrow\,H$ and (2)
pomeron exchange, namely $pp\rightarrow\,\p\,\p\,\rightarrow\,H$,
where $\p$ denotes a pomeron. For $\gamma\,\gamma$ fusion, there is
no hard re-scattering of the photons to fill up the rapidity gaps,
so that large rapidity gaps are automatically present. The
motivation for considering $\p\,\p\rightarrow\,H$, is the
observation that for gluon gluon fusion, namely the process
$gg\rightarrow\,H$ the colour flow induces many secondary parton
showers which fill up the rapidity gaps. Instead the process
$\p\,\p\,\rightarrow\,H$ is a colour singlet exchange, where the
colour
flow is screened, and the large rapidity gaps are preserved.\\

However in the case of $\p\p$ fusion, there are hard re-scattering
corrections,  giving additional inelastic scattering, which will
give emission  filling up the rapidity gaps (see ref. \cite{1}). To
guarantee the presence of large rapidity gaps after scattering in
central exclusive production, in the case of $\p\p$ fusion, one has
to multiply by the survival probability. This is the probability
that large rapidity gaps, between the Higgs boson and the emerging
protons, will be present after scattering.\\

The motive of this paper, was driven by the potentially small value
for the survival probability, namely
$<\,\vert\,S^2\vert\,\,>\,=\,4\times\,10^{-3}$, calculated in ref.
\cite{1} for central exclusive Higgs production via $\p\p$ fusion.
This is an order of magnitude less than previous estimates, namely
$<\,\vert\,S^2\vert\,>\,=\,0.02$ for LHC energies (see ref.
\cite{kmr}). Hence, previous calculations in ref. \cite{kmr} of the
cross section for central exclusive Higgs production via $\p\p$
fusion, which included a factor for the survival probability of
$<\,\vert\,S^2\vert\,>\,=\,0.02$, gave for the exclusive cross
section
$\sigma^{\mbox{exc}}_{\p\p}\Lb\,pp\rightarrow\,p+H+P\Rb\,=\,3\,\mbox{fb}$.
Since the survival probability is predicted in ref. \cite{1} to be
an order of magnitude less, it follows that the cross section
$\sigma^{\mbox{exc}}_{\p\p}\Lb\,pp\rightarrow\,p+H+P\Rb\,$ for
central exclusive Higgs production will also be an order of
magnitude smaller. It also follows that this cross section will be
competitive with the cross section for central exclusive Higgs
production via $\gamma\,\gamma$ fusion, which is predicted in ref.
\cite{kmr} to be
$\sigma^{\mbox{exc}}_{\gamma\gamma}\Lb\,p+p\rightarrow\,p+H+p\Rb\,=\,0.1
\mbox{fb}$.

To illustrate that for the process of \eq{ppH}, the cross sections
will be competitive for $\gamma\gamma$ fusion and $\p\p$fusion, it
is instructive to consider the diagrams for these processes. The
notation used for the couplings in the standard model are the
following.

\begin{center}\begin{tabular}{|c|c|r|}
    \hline
coupling constant & expression &  value ( $\mbox{GeV}\,/\,c^2$ )    \\
    \hline
    & & \\
$\alpha_{em}$   & $\frac{e^2}{4\pi}\,$ & $\approx\,\frac{1}{137}$  \\
& &\\
$G_F$  & $\frac{\sqrt{2}g_w^2}{8m_w^2}\,$ & $1.17\,\times\,10^{-5}$ \\
& &\\
$m_w$  & $\frac{1}{2}v\,g_w$ & 80 \\
& &\\
$M_H$ & $\sqrt{2\lambda\,v^2}$ & 120\\ & &\\
$\as\Lb\,M_H^2\Rb\,$ & $\frac{g_s^2\Lb\,M_H^2\Rb\,}{4\pi} $&  0.12\\
& &\\ \hline
\end{tabular}
\end{center}.\\

where the mass of the Higgs boson, is derived from the vacuum
expectation value of the Higgs $SU\Lb2\Rb\,$ weak isodoublet, which
is
$\begin{pmatrix}\,H^{-}\\H\,\end{pmatrix}_{\mbox{vev}}\,=\,\frac{1}{\sqrt{2}}\begin{pmatrix}0\\v\end{pmatrix}$,
where $v\,=\,\sqrt{-\frac{\mu^2}{\lambda}}$, and $\mu$ and $\lambda$
are
 parametres in the Higgs potential, which is introduced into
the standard model in the spontaneous symmetry breaking of
$SU_{L}\Lb\,2\Rb\,\times\,U_{Y}\Lb\,1\Rb\,\rightarrow\,U_{EM}\Lb\,1\Rb\,
$, which is responsible for giving rise to the W and Z boson masses.
In the case for $\gamma\gamma$ fusion shown in \fig{pe}, there are
four vertices proportional to $\alpha_{em}$ coupling the photons to
the two protons and the photons either side of the subprocess for
$\gamma\gamma\rightarrow\,H$. A factor of
$g_w^2\,=\,4\sqrt{2}G_F\,m_w^2$ should also be included, to account
for the weak coupling of the Higgs to the sub-process, depicted by
the shaded area in \fig{pe}. So it is expected that the cross
section will be proportional to

\beq
\sigma^{\mbox{exc}}_{\gamma\gamma}\Lb\,p+p\rightarrow\,p+H+P\Rb\,\propto\,4\,\sqrt{2}\,G_F\,m_w^2\,\alpha_{em}^4\,=\,0.6\,\mbox{fb}\,\label{em}
\eeq

where units are defined as $ 1\,\mbox{GeV}^{-2}\,=\,0.3893\,\,
\mbox{mb}\,$. In central exclusive Higgs production for the case for
$\gamma\gamma$ fusion, all the couplings of the photons shown in
\fig{pe} are known constants, namely they are proportional to
$\alpha_{em}$. Hence, the cross section for this diagram  can be
calculated exactly. On the other hand, when considering central
exclusive Higgs production for the case for $\p\p$ fusion shown in
\fig{dd}, the couplings of the gluons are not constants. In \fig{dd}
there are four gluon couplings with the protons, giving a
contribution to the cross section proportional to
$\as^4\Lb\,Q^2\Rb\,$, where $Q^2$ is the momentum transferred along
the pomeron, and it is assumed that $\as\Lb\,Q^2\Rb\,\sim\,0.2$.
There are also two gluon couplings with the subprocess for
$\p\p\,\rightarrow\,H$, giving a contribution to the cross section
proportional to $\as^2\Lb\,M_H^2\Rb\,$. Taking the mass of the Higgs
to be $M_{H}\sim{}100GeV$, then it is expected that
$\as\Lb\,M_H^2\Rb\,\sim\,0.12$. Also in the case of $\p\p$ fusion, a
factor of $g_w^2\,=\,4\sqrt{2}G_F\,m_w^2$ should also be included,
to account for the weak coupling of the Higgs to the sub-process
shown in \fig{dd}. For $\p\p\,$ fusion, this subprocess is the quark
triangle subprocess shown in \fig{tr2}. Since the gluon itself
couples weakly to the Higgs boson, only the contribution of the
quark triangle subprocess is taken into account. The amplitude for
the quark triangle subprocess, is derived in \sec{sec:qtr} for the
electromagnetic case, where $\as\,$ replaces $\alpha_{em}$. After
multiplying by a factor for the survival probability, which includes
the survival probability
$<\,\vert\,S^2_{\mbox{hard}}\vert\,>\,=\,4\times\,10^{-3}$ which
takes into account hard re-scattering of the pomeron, and
$<\vert\,S^2_{\mbox{soft}}\vert\,>\,=\,5\times\,10^{-2}$ which takes
into account soft re-scattering of the pomeron, then one obtains for
the exclusive cross section for central exclusive Higgs production,
in the case of $\p\p$ fusion the value

\beq
\sigma^{\mbox{exc}}_{\p\p}\Lb\,p+p\rightarrow\,p+H+p\Rb\,\propto\,4\,\sqrt{2}\,G_F\,m_w^2\,\as^2\Lb\,M_H^2\Rb\,\as^4\Lb\,Q^2\Rb\,<\,\vert\,S^2_{\mbox{hard}}\vert\,><\vert\,S^2_{\mbox{soft}}\vert\,>\,=\,0.9\,\mbox{fb}\label{pp}\eeq

Comparing the estimates of \eq{em} and \eq{pp}, it is expected that
$\sigma^{\mbox{exc}}_{\gamma\gamma}\Lb\,p+p\rightarrow\,p+H+p\Rb\,$
and $\sigma^{\mbox{exc}}_{\p\p}\Lb\,p+p\rightarrow\,p+H+p\Rb\,$ will
be competitive. This is the motivation for this paper which
re-examines
$\sigma^{\mbox{exc}}_{\gamma\gamma}\Lb\,p+p\rightarrow\,p+H+p\Rb\,$
for electromagnetic Higgs production.\\

This paper is organised in the following way. In \sec{sec:dd}, the
details of the calculation of the cross section for central
exclusive Higgs production in the case for $\p\p$ fusion is
explained in detail. The cross section which is obtained, is
multiplied by the factor for the survival probability in ref.
\cite{1}, to give the exclusive cross section. In \sec{sec:photon},
the cross section for central exclusive Higgs production, in the
case of $\gamma\gamma$ fusion is calculated. The mechanism
$\gamma\gamma\,\rightarrow\,H$ proceeds via the fermion triangle and
Boson loop sub-processes illustrated in \fig{leptr} and \fig{BLG}.
The total cross section for central exclusive Higgs production, in
the case of $\gamma\gamma$ fusion is calculated by taking the sum
over all the contributions for these sub-processes for the
$\gamma\gamma\rightarrow\,H$ mechanism. Finally, in the conclusion
the results found in
\sec{sec:dd} and \sec{sec:photon}, are compared.\\

All calculations in this paper, are  based on the Feynman rules for
the standard electro-weak theory, given in \fig{feyn}, which is to
be found in the appendix.

\section{Double diffractive Higgs production at the LHC}
\label{sec:dd}

\FIGURE[h]{ \centerline { \epsfig{file=
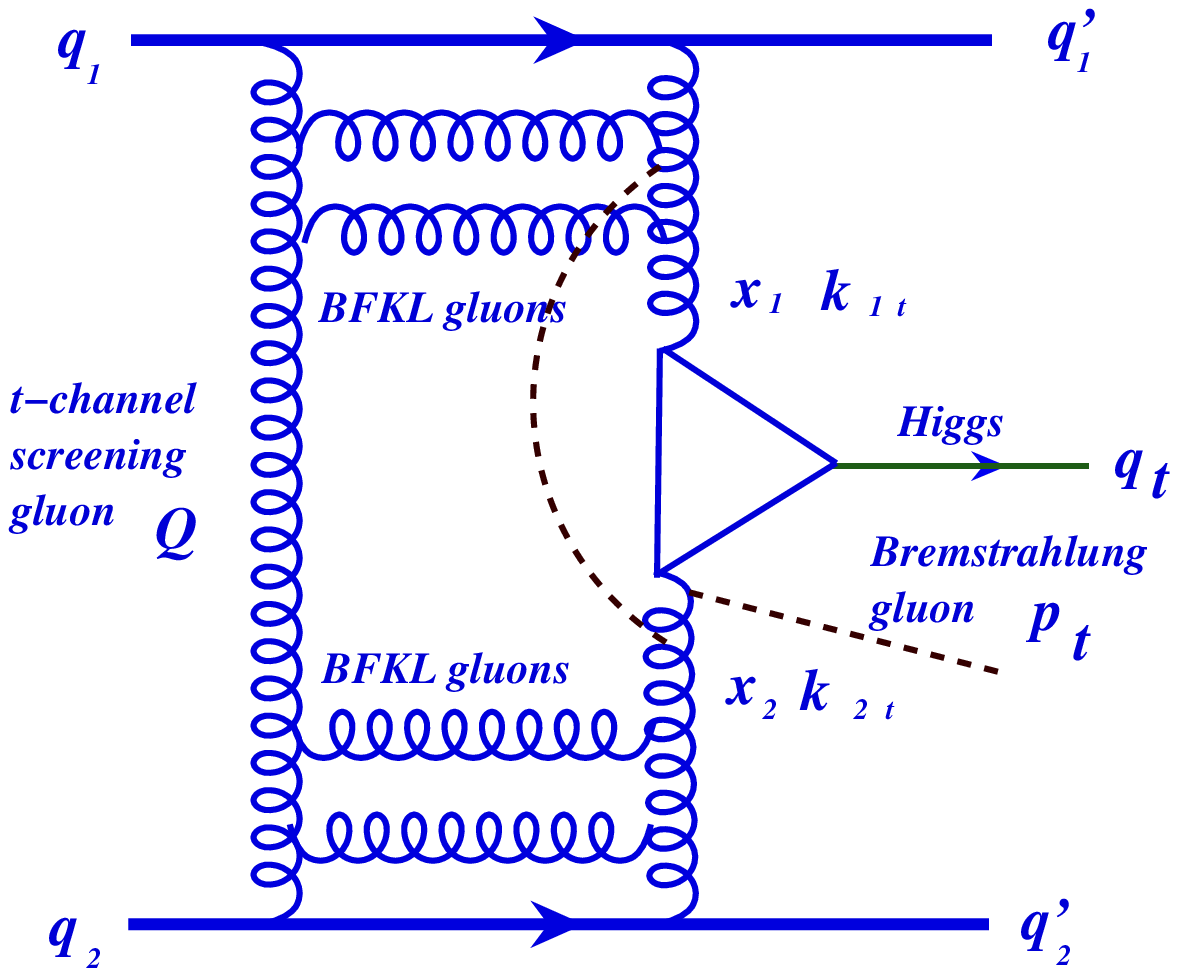,width=133mm,height=80mm}} \caption{Double diffractive
Higgs production in the Born approximation} \label{dd} }

In this section, an explicit expression is derived for the Born
amplitude for double diffractive Higgs production (see \fig{dd}).
The Higgs couples weakly to the gluon, so the main contribution
comes from the quark triangle subprocess (see \fig{tr2}), and all
six flavours of quarks are taken into account. The second $t$
channel gluon in \fig{dd} is included. This is because of the large
rapidity (LRG) gap between the Higgs and the proton, which demands
that in the $t$ channel, one has colorless exchange. Indeed, if the
LRG wasn't present between the protons, then the Higgs could simply
be produced from gluon gluon fusion in a single channel. However,
the colour flow induced by a single channel exchange process, could
produce many secondary particles. These secondary particles could
fill up the LRG. To screen the colour flow, it is necessary to
exchange a second $t$ channel gluon. At lowest order in
$\alpha_{s}$, this gluon couples only to the incoming quark lines.
The Born amplitude for double diffractive Higgs production by gluon
exchange, is given by the expression \cite{4,6}

\bea
\frac{4}{9}\frac{2s}{M_{H}^{2}}A\,\vec{k}_{1}\cdot{}\vec{k}_{2}\,\int{}\frac{d^{2}Q}{Q^{2}}\frac{\vec{k}_{1T}\cdot{}\vec{k}_{2T}}{k_{1}^{2}k_{2}^{2}}8\alpha{}_{s}^{2}\left(Q^{2}\right)\,=\,\frac{4}{9}\frac{s}{M_{H}^{2}}A\frac{M_{H}^{2}}{2}\int{}\frac{d^{2}Q}{Q^{2}}\frac{\vec{k}_{1T}\cdot{}\vec{k}_{2T}}{k_{1}^{2}k_{2}^{2}}8\alpha{}_{s}^{2}\left(Q^{2}\right)\label{E:2.1}
\eea
\\

where $\vec{k}_{1}\cdot{}\vec{k}_{2}=\frac{M_{H}^{2}}{2}$ has been
used. In \eq{E:2.1}, the  Weizs\"{a}cker - Williams approach,
explained in ref. \cite{5}
 has been used. The factor
$A\,\vec{k}_{1}\cdot{}\vec{k}_{2}\,$ is the amplitude for the quark
triangle subprocess of \fig{tr2}, where $A$ takes the value,
\cite{Rizz,2,Higgs,daw}

\FIGURE[h]{ \centerline { \epsfig{file=
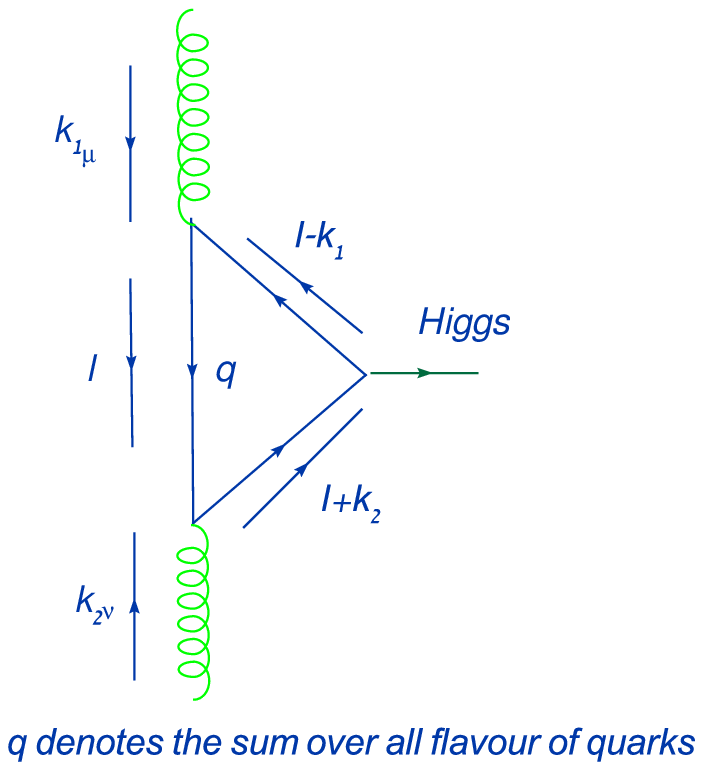,width=100mm,height=100mm}} \caption{Quark triangle
subprocess for Higgs production in gluon gluon fusion} \label{tr2} }

\begin{align}
A=\frac{2}{3}\left(\frac{-\alpha{}_{s}\left(M_{H}^{2}\right)\left(\sqrt{2}G_{F}\right)^{\frac{1}{2}}}{\pi{}}\right)\label{E:A}
\end{align}

The Born amplitude shown in \fig{dd}, is extended to proton
scattering instead of just quark scattering here.
 The typical momentum transferred,
$t_1=\Lb\,q_1-Q\Rb\,^2$ and $t_2=\Lb\,q_2-Q\Rb\,^2$, are rather
small and of the order of $\frac{1}{b}$, where $b$ is the slope of
the gluon - proton form factor, and can be estimated to be
$b=5.5GeV^2$. Therefore, one takes into account the proton couplings
to the gluon ladder (see \fig{dd}), by including the proton form
factors in the gaussian form
\bea\,\exp\Lb\,-\frac{1}{2}bt_1-\frac{1}{2}bt_2\Rb\,&
&\mbox{where}\,\,\,\,\,\,t_1=\Lb\,q_1-Q\Rb^2\,\,\,\,\,\mbox{and}\,\,\,\,\,\,t_2=\Lb\,q_2-Q\Rb^2\label{E:ff}\eea

to describe the dependence on the transferred momentum. If the
momentum transfer is small, it can be assumed that
$k_{1}\sim{}k_{2}\sim{}Q$. Therefore, with the definition of
\eq{E:ff}, $t_{1}\rightarrow{}0$ and $t_{2}\rightarrow{}0$. Hence,
the proton form factors $e^{-\frac{1}{2}\,bt_1}$ and
$e^{-\frac{1}{2}\,bt_{2}}$ tend to 1, and can be taken outside the
integral, and the Born amplitude behaves as \beq
\frac{2}{9}\,A\,e^{-\frac{1}{2}\,bt_{1}}e^{-\frac{1}{2}\,bt_{2}}\int{}\frac{d^{2}Q}{Q_{\bot}^{4}}\,8\as\Lb\,Q^2\Rb\,\label{E:proton}\eeq

To consider the exclusive process only, with the condition of the
LRG, bremsstrahlung gluons must be suppressed. The bremsstrahlung
gluons are shown by the dashed lines in \fig{dd}, which
 are suppressed by multiplying by the Sudakov form factor
\begin{equation}
F_{s}=e^{-S\left(k_{\bot}^{2},E_{\bot}^{2}\right)}\label{E:2.2a}
\end{equation}

$F_{s}$ is the probability not to emit bremsstrahlung gluons. $S$ is
the mean multiplicity of Bremsstrahlung gluons given as
\begin{equation}
S\left(k_{\bot}^{2},E_{\bot}^{2}\right)=\int^{E_{\bot}^{2}}_{k_{\bot}^{2}}\frac{dp_{\bot}^{2}}{p_{\bot}^{2}}\int^{\frac{M_{H}}{2}}_{p_{\bot}}\frac{d\omega{}}{\omega{}}\frac{3\alpha{}_{s}\left(p_{\bot}^{2}\right)}{\pi{}}\left(\ln{}\left(\frac{E_{\bot}^{2}}{k_{\bot}^{2}}\right)\right)^{2}\label{E:2.2b}
\end{equation}

Secondly, evolution of BFKL ladder gluons  between the two channels
(see \fig{dd}), must be taken into account. For proton scattering
instead of quark scattering, the naive coupling of the gluons to the
external quarks must be replaced by a coupling of the gluons to the
external proton lines. To include both of these modifications, the
naive gluon density for quarks is replaced by the density for
protons  by the following substitution

\begin{equation}
\frac{4\alpha{}_{s}\left(Q^{2}\right)}{3\pi{}}\rightarrow{}f\left(x,Q^{2}\right)\label{E:2.3}
\end{equation}

where $f\left(x,k^{2}\right)$ is the un-integrated gluon density of
the proton. After including the Sudakov form factor of \eq{E:2.1},
and the gluon density function of \eq{E:2.3}, the amplitude in
\eq{E:2.1} becomes

\bea
\mbox{M}_{\p\p}\left(\mbox{p+p}\rightarrow{}\mbox{p+H+p}\right)&=&A\pi{}^{3}s\int{}\frac{dQ_{\bot}^{2}}{Q_{\bot}^{4}}e^{-S\left(k_{\bot}^{2},E_{\bot}^{2}\right)}f\left(x_{1},Q^{2}\right)f\left(x_{2},Q^{2}\right)\label{E:2.4}\\
\mbox{where for the gluon
densities}\,\,\,\,\,\,\,f\Lb\,x_{1,2}\Rb\,&=&2\Lb\,Q^2\Rb\,^{\gamma_{1,2}}e^{\omega\Lb\,\gamma_{1,2}\Rb\,\ln\Lb\,\frac{1}{x_{1,2}}\Rb\,}\label{f}
\eea

where $\gamma_{1,2}$ are the anomalous dimensions and the numerical
coefficient $2$ in \eq{f} can be taken from MRST-2002-NLO
parameterizations. Using \eq{f}, the integration over $Q_{\bot}$ and
over $\gamma_1$ and $\gamma_2$ can be evaluated. It turns out that
the integrand of \eq{E:2.4} has a saddle point given by
$\ln\,Q_{\bot}^2=\ln\,\frac{M_H^2}{4}+\frac{2\pi}{3\as}\Lb\,\gamma_1+\gamma_2-1\Rb\,$,
and
 the essential values of $\gamma_1$ and $\gamma_2$ are close to
$\frac{1}{2}$. Hence, the typical $Q_{\bot}$ is rather large, and
depends on the mass of the Higgs. After integrating over $\gamma_1$
and $\gamma_2$ and $q_{\bot}$, the final result for the amplitude
$\mbox{M}_{\p\p}\Lb\,\mbox{p+p}\rightarrow\,\mbox{p+H+p}\Rb\,$ is
derived in the appendix (see \sec{sec:M}), and the final result is
given in \eq{E:B11} as

\begin{align}
\mbox{M}_{\p\p}\left(\mbox{p+p}\rightarrow{}\mbox{p+H+p}\right)=&-2A\pi^{4}s\Lb\frac{4\pi^2}{3\alpha_s}\Rb^{\frac{1}{2}}\frac{\exp\Lb-\frac{\Lb\ln\,\frac{M_{H}^{2}}{4}\Rb^2}{\omega\,"\Lb\frac{1}{2}\Rb\ln\,s_{1}}\Lb\frac{1}{2}-\frac{\frac{\pi}{3\alpha_{s}}}{\omega\,"\Lb\frac{1}{2}\Rb\ln\,s_{1}}\Rb
\Rb}{\Lb\frac{2\pi}{3\alpha_{s}}+\ln\,\frac{M_{H}^{2}}{4}+\omega\,"\Lb\frac{1}{2}\Rb\ln\,s_{1}\Rb}\label{E:2.5}
\end{align}

Now that the amplitude is known,
$\sigma_{\p\p}\Lb\,\mbox{p+p}\rightarrow\,\mbox{p+H+p}\Rb\,$ for
central Higgs production can be calculated for the case of $\p\p$
fusion. To derive the cross section for exclusive central Higgs
production, one has to multiply
 by a factor which takes into account the survival probability for large
rapidity gaps $<{}|{}S^{2}_{\mbox{hard}}|{}>{}=0.004$, to suppress
hard re-scattering. Therefore, the cross section
$\sigma^{\mbox{exc}}_{\p\p}\Lb\,\mbox{p+p}\rightarrow\,\mbox{p+H+p}\Rb\,$,
for central exclusive Higgs production, without any further hard
re-scattering for the case of $\p\p$ fusion, takes the value

\begin{align}
<{}|{}S^{2}_{\mbox{hard}}|{}>{}\sigma_{\p\p}\Lb\,\mbox{p+p}\rightarrow\,\mbox{p+H+p}\Rb\,\,=\,\,\,\sigma^{\mbox{exc}}_{\p\p}\Lb\,\mbox{p+p}\rightarrow\,\mbox{p+H+p}\Rb\,\,=\,\,\,0.47\,\mbox{fb}\label{E:d96GGG}
\end{align}

\section{Electromagnetic Higgs production at the LHC}
\label{sec:photon}

In the case of central exclusive Higgs production for the case of
$\gamma\gamma$ fusion, shown in \fig{pe}, there is no hard
re-scattering to take into account, and all the couplings are known
precisely. The shaded area in \fig{pe} depicts the subprocess for
the mechanism $\gamma\gamma\,\rightarrow\,H$. The possible
mechanisms are illustrated in \fig{leptr} and \fig{BLG}, and their
contributions to the amplitude are calculated in this section. Gauge
invariance requires that the contribution of a subprocess for the
mechanism $\gamma\gamma\,\rightarrow\,H$, takes the form

\FIGURE[h]{ \centerline{ \epsfig{file=
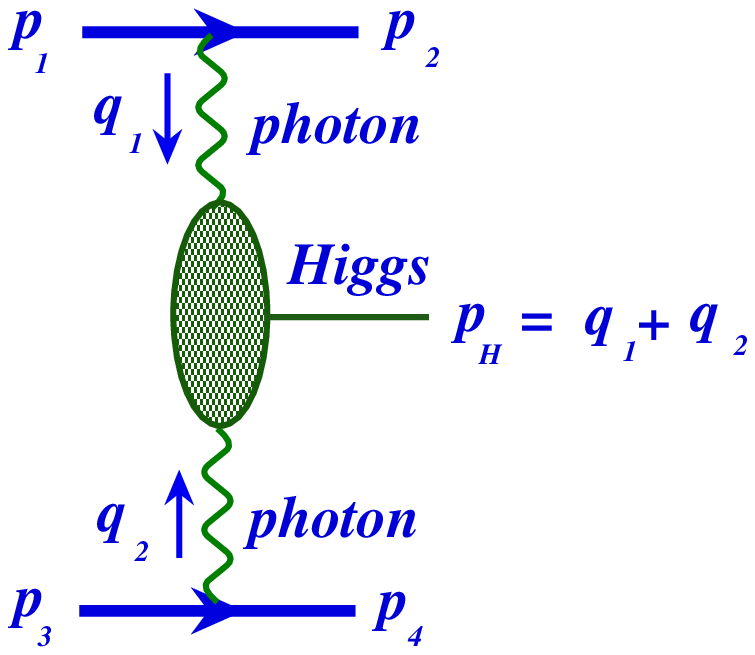,width=90mm,height=60mm}} \caption{Diffractive Higgs
production in single channel photon exchange.} \label{pe} }

\beq\,A^{\mn}\,=\,A\Lb\,q_1^{\nu}q_2^{\mu}\,-\,\frac{M_H^2}{2}g^{\mn}\Rb\,\label{gsp}\eeq

where $A$ is a constant, depending on the particular subprocess. It
turns out that for the case of when the subprocess is the fermion
triangle shown in \fig{leptr}, summed over all six flavours of
quarks, and all three lepton flavours, then the expression for the
amplitude takes the form of \eq{gsp}. However in the case of when
the subprocess is one of the Boson loops shown in \fig{BLG}, the
expression for the amplitude is not of the form of \eq{gsp}. The
correct statement is that the sum over all the amplitudes for the
sub-processes shown in \fig{BLG}, gives a gauge invariant expression
of the form of \eq{gsp}.

The amplitude for the diagram of \fig{pe}, where the process
$\gamma\gamma\rightarrow\,H$ proceeds via the sum over the fermion
triangle shown in \fig{leptr}, and Boson loops shown in \fig{BLG},
is given by

\beq
\mbox{M}_{\gamma\gamma}\Lb\,\mbox{p+p}\rightarrow\,\mbox{p+H+p}\Rb\,=\,-\frac{4\pi\alpha_{em}}{q_{1}^{2}q_{2}^{2}}\,4p_{2}^{\mu}p_{1}^{\nu}\Lb\,A_{\mn}^{f}+A_{\mu\nu}^{b}\Rb\,\label{pr}
\eeq

where
$A^f_{\mn}\,=\,A_f\Lb\,q_1^{\nu}q_2^{\mu}-\frac{M_H^2}{2}g^{\mn}\Rb\,
$ denotes the amplitude of the quark/anti-quark triangle subprocess,
summed over all six quark flavours/anti-quark flavours
($q,/\bar{q}\,=\,u/\bar{u},d/\bar{d},s/\bar{s},c/\bar{c},t/\bar{t},b/\bar{b}$)
and all three lepton/anti-lepton flavours
($L^{\pm}\,=\,e^{\pm},\mu^{\pm},\tau^{\pm}$), shown in \fig{leptr},
and
$A^b_{\mn}\,=\,A_b\Lb\,q_1^{\nu}q_2^{\mu}-\frac{M_H^2}{2}g^{\mn}\Rb\,
$ denotes the amplitude of the sum over all the Boson loop
sub-processes shown in \fig{BLG}.  At this point, the
\emph{Weizsacker - Williams} formula is used, as explained in refs.
\cite{4,5,6}. In this approach, the substitution
$p^{\mu}_1p^{\nu}_2A_{\mu\nu}\,=\,-\frac{2S}{M_{H}^2}q^{\mu}_{1\bot}q^{\nu}_{2\bot}A_{\mu\nu}$
is used. In the notation used in this paper, $q^{\mu}_{1\bot}$ and
$q^{\mu}_{2\bot}$ denotes two dimensional vectors, in the plane
transverse to the direction of the momenta of the two incoming
protons $p^{\mu}_1$ and $p^{\nu}_2$. Hence, the amplitude of \eq{pr}
can be written as

\begin{align}
\mbox{M}_{\gamma\gamma}\Lb\,\mbox{p+p}\rightarrow\,\mbox{p+H+p}\Rb\,=-\frac{4\pi\alpha_{em}}{q_{1}^{2}q_{2}^{2}}\frac{2s}{M_{H}^{2}}\,4q_{2\bot}^{\mu}q_{1\bot}^{\nu}\Lb\,A_{\mn}^{f}+A_{\mn}^b\Rb\,
\label{E:y}
\end{align}

In order to calculate the cross section, one has to integrate the
squared amplitude over all the transverse momenta $q_{1\bot}$ and
$q_{2\bot}$. For central exclusive Higgs production in the case of
$\gamma\gamma$ fusion, it is required that the lower limits of
integration  are
$\left(q_{1\bot}^{min}\right)^{2},\left(q_{2\bot}^{min}\right)^{2}\,=\,m_{p}^{2}\sqrt{\frac{M_{H}^{2}}{s}}$,
where $m_p$ is the proton mass, which is assumed in this paper to be
$1$ GeV. The upper limits of integration, are taken from the
electromagnetic form factors for the proton, namely
$G_p\Lb\,q^2\Rb\,=\frac{1}{2\Lb\,1+\frac{q^2}{0.72}\Rb\,}$, from
which the upper limits of the integration are derived to be
$\Lb\,q_{1\bot}^{max}\Rb^2\,=\,\Lb\,q_{1\bot}^{max}\Rb^2\,=\,0.72$.

\begin{align}
\sigma_{\gamma\gamma}\Lb\,\mbox{p+p}\rightarrow\,\mbox{p+H+p}\Rb\,=\,\frac{\alpha_{em}^2\,}{32\pi}\ln\frac{s}{m^2}\Lb\,A_f+A_b\Rb^2\ln^2\,\Lb\,\frac{0.72}{m_p^2\sqrt{\frac{M_H^2}{s}}}\Rb\,\label{E:35aa}
\end{align}

where $y\,=\,\ln\frac{s}{m^2}$ is the rapidity gap between the two
incoming protons in \fig{pe}, and $m$ is the proton mass, assumed to
be $1$ GeV. This calculation is for central exclusive Higgs
production at the LHC, where it is expected that $\sqrt{s}\,=\,14000
\,\mbox{GeV}$, which gives for the value of the rapidity gap
$y\,=\,19$.

\subsection{The fermion triangle subprocess for Higgs production in $\gamma\gamma$ fusion}
\label{sec:qtr} \setcounter{equation}{0}
\numberwithin{equation}{subsection}

Central exclusive Higgs production for the case of $\gamma\gamma$
fusion, can proceed through the subprocess
$\gamma\gamma\,\rightarrow\,\mbox{fermion triangle}\,\rightarrow\,H$
shown in \fig{leptr}, where the fermions include the six flavours of
quarks and anti quarks ($u\,,\,d\,,\,s\,,\,c\,,\,t\,,\,b$) and the
three lepton and anti lepton flavours $(e\,,\,\mu\,,\,\tau\,)$. A
derivation of the amplitude for the fermion triangle, can be found
in refs. \cite{Rizz,daw,22} for the case where the mass of the
fermion in the triangle is much larger than the Higgs mass. In this
section, the amplitude of the fermion triangle is derived by taking
the sum over all the fermions which could contribute, including the
six quark flavours and the three lepton flavours. The
$H\,\rightarrow\,f\,f$ vertex coupling, is proportional to the mass
$m_f$ of the fermion at the vertex, so that the sub-process
amplitude of the fermion triangle will be proportional to the mass
of the fermion in the triangle. The fermion masses are assumed to
take the following values listed in the table below.

\begin{center}\begin{tabular}{|c|r|}
    \hline
fermion &  mass ( $\mbox{GeV}\,/\,c^2$ )    \\
    \hline
    quarks & \\
    \hline
u   &  $3\,\times\,10^{-3}$ \\
d & $6\,\times\,10^{-3}$\\
s & $1.3\,$\\
c & $0.1$ \\
t & $175$ \\
b & $4.3$ \\

\hline
fermions & \\
\hline
e & $5.11\,\times\,10^{-4}$\\
$\mu$ & $0.106$ \\
$\tau$ & $1.7771$ \\ \hline
\end{tabular}
\end{center}.\\

Hence, these values indicate that the most significant contribution
to the amplitude will come from the top quark triangle.

\FIGURE[h]{ \centerline{\epsfig{file=
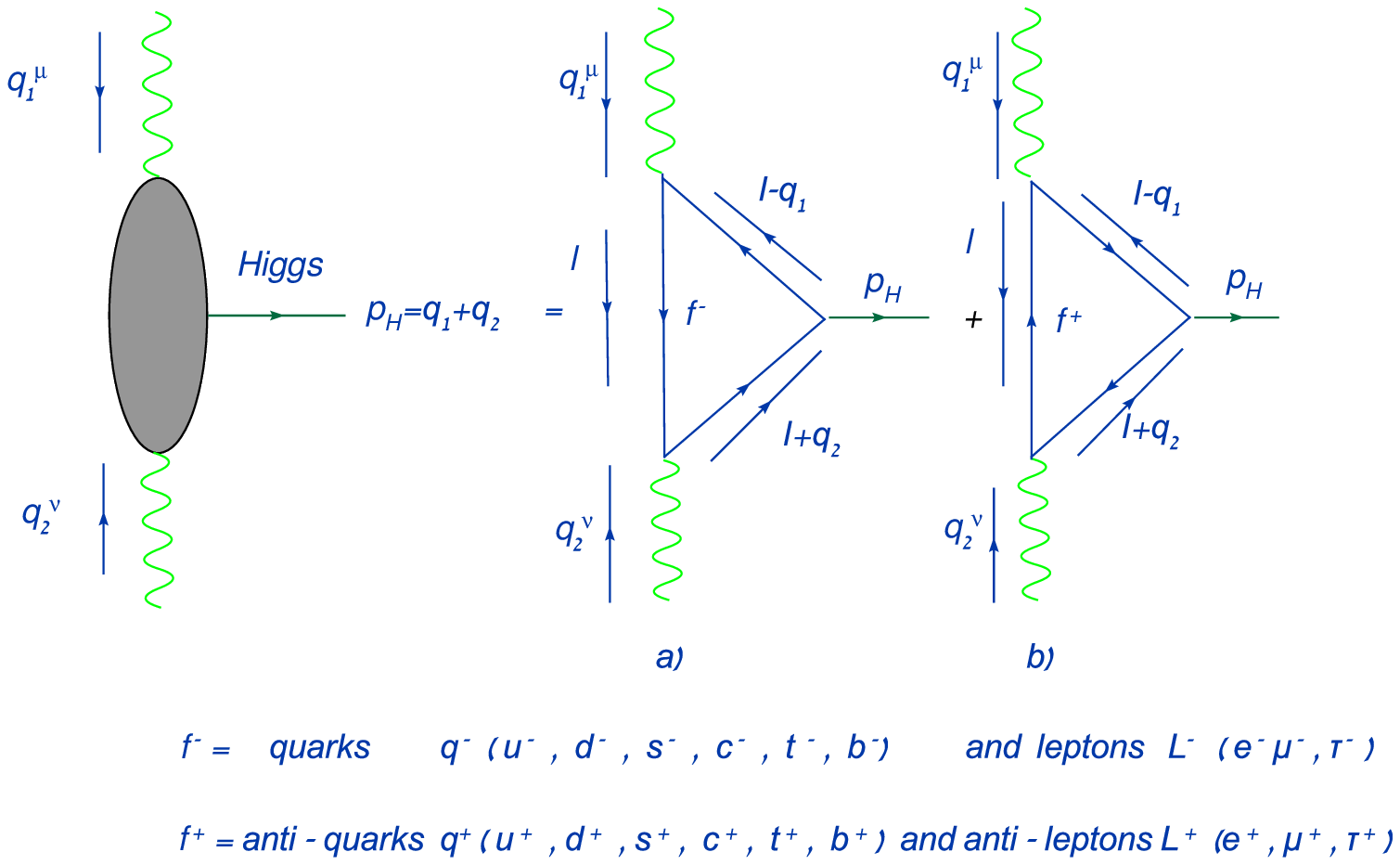,width=158mm,height=95mm}} \caption{Fermion triangle
subprocess for Higgs production through $\gamma\gamma$ fusion}
\label{leptr}}

To derive explicitly the amplitude of the subprocess  \fig{leptr},
the labeling of momenta shown in the diagram is used, and the
following notation is introduced. \bea
D_{1}=l^{2}-m_{f}^{2}\,\,\,\,\,\,\,\,\,\,\,\,\,\,\,\,D_{2}=\left(l-q_{1}\right)^{2}-m_{f}^{2}
\,\,\,\,\,\,\,\,\,\,\,\,\,\,\,\,\,D_{3}=\left(l+q_{2}\right)^{2}-m_{f}^{2}\label{E:D}
\eea

where $m_f$ denotes the mass of the fermion which forms the
triangle, which could be one of the quark flavours or one of the
lepton flavours. Then the amplitude for the fermion triangle
subprocess, summed over all possibilities of quark and anti-quark
flavours, and lepton and anti-lepton flavours takes the form

\bea A^{\mn}_q\,&=&\,
\sum_f\,4\pi{}\alpha_{em}\left(\sqrt{2}G_{F}\right)^{\frac{1}{2}}m_{f}\int{}\frac{d^{d}l}{\left(2\pi{}\right)^{d}}\left(\frac{I^{\mn}_f}{D_{1}D_{2}D_{3}}\right)\,\,\,\,\,\mbox{where}\,\,\,\sum_{f}=\,\sum_{q}\,+\sum_{L\,=\,e\,,\,\mu\,,\,\tau}\label{cf}\eea

where $\sum_q$ denotes the sum over all six quark flavours
$q\,=\,u\,,\,d\,,\,s\,,\,c\,,\,t\,,\,b$ and
$\sum_{L\,=\,e\,,\,\mu\,,\,\tau}$ denotes the sum over all three
lepton flavours $L\,=\,e\,,\,\mu\,,\,\tau$. On the RHS of \eq{cf},
$d$ is the space-time dimension, and at the end f the calculation,
$d\,\rightarrow\,4$ is imposed. The reason for not specifying this
in the beginning, is because it will be necessary to use dimensional
regularisation to cancel divergences, which requires integration
over $d\,+\,\epsilon$ dimensions, in the limit that
$d\,\rightarrow\,4$ and $\epsilon\,\rightarrow\,0$. Using the
notation shown in the diagram of \fig{leptr} for the flow of momenta
in the quark and anti - quark triangles, the trace term $I^{\mn}_f$
is given by \bea
I_{f}^{\mn}\,&=&\,Tr\left(\gamma{}^{\rho{}}\left(l_{\rho}+q^{2}_{\rho}+m_{f}\right)\gamma^{\mu}\gamma^{\sigma}\left(l_{\sigma}+m_{f}\gamma^{\nu}\gamma^{\tau}\left(l_{\tau}-q^{1}_{\tau}+m_{f}\right)\right)\right)\notag\\
&\,&\,+
Tr\left(\gamma{}^{\rho{}}\left(-l_{\rho}+q^{1}_{\rho}+m_{f}\right)\gamma^{\nu}\gamma^{\sigma}\left(-l_{\sigma}+m_{f}\gamma^{\mu}\gamma^{\tau}\left(-l_{\tau}-q^{2}_{\tau}+m_{f}\right)\right)\right)\notag\\
&\,=\,&\,8m_{f}\left(q_{1}^{\nu}q_{2}^{\mu}+4l^{\mu}l^{\nu}+2\left(l^{\mu}q_{1}^{\nu}-l^{\nu}q_{2}^{\mu}\right)-g^{\mu\nu}\left(\vec{q}_{1}\cdot{}\vec{q}_{2}+l^{2}-m_{f}^{2}\right)\right)\label{trace}
\eea

where $\vec{q}_1$ and $\vec{q}_2$ denotes four dimensional vectors.
The first line on the RHS of \eq{trace} corresponds to the
contribution given by the triangle formed by the fermions, ( i.e.
quarks $q\,=\,(u\,,\,d\,,\,s\,,\,c\,,\,t\,,\,b\,)$ and negatively
charged leptons
$L^{-}\,=\,\Lb\,\,e^{-}\,,\,\mu^{-}\,,\,\tau^{-}\Rb\,$, and the
second line corresponds to the contribution given by the triangle
formed by the anti - fermions ( i.e. anti - quarks
$\bar{q}\,=\,(\bar{u}\,,\,\bar{d}\,,\,\bar{s}\,,\,\bar{c}\,,\,\bar{t}\,,\,\bar{b}\,$)
and positively charged leptons
$L^{+}\,=\,\Lb\,\,e^{+}\,,\,\mu^{+}\,,\,\tau^{+}\Rb\,$ ).
Introducing Feynman parametres to rewrite the quotient
$\Lb\,D_1\,D_2\,D_3\Rb^{-1}$ in a more convenient form, \eq{cf}
simplifies to

\bea A^{\mu\nu}_f\,
&=&\,2\sum_f4\pi{}\alpha{}_{em}\left(G_{F}\sqrt{2}\right)^{\frac{1}{2}}\int{}\frac{d^{d}\tilde{l}}{\left(2\pi{}\right)^{d}}\int^{1}_{0}dx\,\int^{1-x}_{0}dy\,\frac{8m_{f}I^{\mu\nu}_{f}}{\left(\tilde{l}^{2}-\Delta_f\Lb\,x,y\Rb\,\right)^{3}}\notag\\
\mbox{where}\,\,\,\,\,\,\Delta_f\,&=&\,m_f^2\,-\,M_H^2\,x\,y\notag\\
\mbox{and}\,\,\,\,\,\,\,\,\tilde{l}^{\mu}\,&=&\,l^{\mu}\,-x\,q_1^{\nu}\,+y\,q_2^{\mu}\label{cf2}\eea

Note in the last step, it was assumed that
$q_{1}^{2}\ll{}q_{2}^{2}\ll{}M_{H}^{2},m_{f}^{2}$ so that these
terms can be ignored. From the kinematics shown in the diagram of
\fig{pe}, and \fig{leptr}, it is clear that
$2\vec{q}_1\cdot\vec{q}_2\,=\,M_H^2$. The trace term $I_f^{\mn}$ was
given in \eq{trace} in terms of the unknown momentum $l^{\mu}$ in
the fermion triangle in \fig{leptr}. In terms of the new variable
$\tilde{l}^{\mu}$, the trace term takes the form

\bea I_{f}^{\mu\nu}
=\,m_f\Lb\,q_{1}^{\nu}q_{2}^{\mu}+4\tilde{l}^{\mu}\tilde{l}^{\nu}-4q_{1}^{\nu}q_{2}^{\mu}x\,y-g^{\mu\nu}\left(\vec{q}_{1}\cdot{}\vec{q}_{2}\left(1-2x\,y\right)-m_{f}^{2}+\tilde{l}^{2}\right)\Rb\,\label{cftrace}
\eea

The details of the integration over the momentum $\tilde{l}$ on the
RHS of \eq{cf2} are given in \sec{sec:Elq} of the appendix, (see
\eq{cf2a} - \eq{dr3}). Here, dimensional regularisation is used, a
technique where one integrates over $d+\epsilon$ dimensions, and
afterwards $d\rightarrow\,4$ and $\epsilon\rightarrow\,0$. This
removes non gauge invariant terms, in the numerator of the integrand
on the RHS of \eq{cf2}. In this way, one obtains the following gauge
invariant result for the RHS of \eq{cf2}.

\bea A^{\mu\nu}_f
&=&\,-\frac{2\,\alpha{}_{em}G_{F}^{\frac{1}{2}}2^{\frac{1}{4}}}{\pi}\,\Lb\,q_1^{\nu}q_2^{\mu}-\frac{M_H^2}{2}g^{\mn}\Rb\,\sum_f\,I_f\label{dr3mt}\\
\mbox{where}\,\,\,\,\,\,\sum_f\,I_f\,&=&\sum_{f\,=\,\mbox{$u\,,\,d\,,\,s\,,\,c\,,\,t\,,\,b\,,\,e\,,\,\mu\,,\,\tau$}}\,m_f^2\,\int^1_0dx\int^{1-x}_0\,dy\frac{1-4xy}{\Delta_f\Lb\,x,y\Rb}\,\,\,\,\notag\\
\,\,\,\,\,\mbox{where}\,\,\,\,\,\Delta_f\Lb\,x,y\Rb\,&=&\,m_f^2-M_H^2xy
\label{intmt}\eea

The integral $I_f$ is evaluated in \sec{sec:Elq} of the appendix
(see \eq{int} - \eq{IR} ).  It turns out that, due to the dependence
of the factor for the $H\,\rightarrow\,f\,f$ vertex coupling on the
mass of the fermion $m_f$, that the only fermion triangle which
gives a significant contribution to the amplitude is for the case
where $m_f\gg\,M_H$. From the table, this is true only for the top
quark $m_t$. Hence, it turns out that the only fermion triangle that
is necessary to take into, is the top quark triangle, and the
contributions from the rest of the triangle sub-processes formed by
the rest of the quarks, and the leptons can be neglected. Using this
result, the amplitude for the contribution of the fermion triangle
has the expression

\begin{align}
A^{\mu\nu}_f\,=A_f\,\left(q_{1}^{\nu}q_{2}^{\mu}-g^{\mu\nu}(\vec{q}_{1}\cdot{}\vec{q}_{2})\right)\,\,\,\,\,\mbox{where}\,\,\,\,A_f\,=\,-\frac{2}{3}\frac{G_F^{\frac{1}{2}}\,2^{\frac{1}{4}}\alpha_{em}}{\pi}
\label{E:900a}
\end{align}

\subsection{Boson loop sub-processes for Higgs production in
$\gamma\gamma$ fusion}

For central exclusive Higgs production, the Higgs can also be
produced through the subprocess
$\gamma\gamma\,\rightarrow\mbox{boson loop}\,\rightarrow\,H$, where
the possible boson loops are shown in \fig{BLG} ( taken from ref.
\cite{2}). $H^{-}$ in \fig{BLG} is an un-physical, charged Higgs
boson, and $\phi^{\pm}$ is a Fadeev-Popov ghost. The formalism for
calculating the amplitude of each diagram in \fig{BLG}, is similar
to the approach used to calculate the amplitude of the fermion
triangle above, in \sec{sec:qtr}. Similarly here, after integration
over the unknown momentum $l$ in the loop of each diagram in
\fig{BLG}, and after integration over Feynman parametres, the
expression for each diagram takes the general form \cite{2}

\bea
A_b^{\mn}\,&=&\,\frac{\alpha_{em}2^{\frac{1}{4}}G_F^{\frac{1}{2}}}{2\pi}\,m_w^2\,\Lb\,\,B\,\Gamma\Lb\,2-\frac{d}{2}\Rb\,g^{\mn}\,+\,C\frac{q_1^{\nu}q_2^{\mu}}{m_w^2}+\,D\frac{M_H^2}{2m_w^2}g^{\mn}\,\Rb\,\label{BSL}
\eea

where $d$ is the dimension of space-time. Terms proportional to
$q_1^2$ and $q_2^2$ were assumed to vanish, and from the kinematics
shown in \fig{pe}, it was assumed that
$\vec{q}_1\,\cdot\,\vec{q}_2\,=\,\frac{M_H^2}{2}\,$. In the limit
that $d\,\rightarrow\,4$, the term proportional to
$\Gamma\Lb\,2-\frac{d}{2}\Rb\,$ on the RHS of \eq{BSL} tends to
infinity. However, when one sums over the contributions to the
amplitude given by all the boson sub-processes shown in \fig{BLG},
these divergencies cancel exactly (see table below). One requires
also, that this sum over all the amplitudes for the boson loop
sub-processes shown in \fig{BLG}, satisfies the condition

\beq \sum\,C\,=\,-\sum\,D\label{gc}\eeq such that the amplitude for
the sum is gauge invariant. In ref. \cite{2}, this sum was taken and
the result was an almost gauge invariant expression, since terms
proportional to $\frac{m_w^2}{M_H^2}$ and higher were neglected. In
the calculation which lead to the results in this paper, the result
gives an exactly gauge invariant expression, after using dimensional
regularisation to remove terms which do not satisfy the gauge
invariance condition.
\\

\begin{center}\begin{tabular}{|c|c|c|r|}
    \hline
graph&  $B$    &   $C$ & $D$ \\
    \hline
a + crossed  &$3\,\Lb\,d-1\Rb\,$ & $-4$ & $5$ \\
b + crossed  & $-2\,\Lb\,d-1\Rb\,$ & $0$ & $2\frac{m_w^2}{M_H^2}$\\
c + d + crossed  & $-\frac{1}{2}\Lb\,d-1\Rb\,$ & $-4$& $2$\\
e + crossed & $0$ & $0$ & $-\frac{m_w^2}{M_H^2}$\\
f + crossed & $0$ & $0$ & $1$\\
g + h + crossed & $1$ & $0$ & $0$\\
i + crossed & $-2$ & $0$ & $0$\\
2j + crossed & $-\frac{1}{2}$ & $0$ & $-\frac{m_w^2}{M_H^2}$\\
 \hline
 sum & $\frac{1}{2}\,d\,-2\,$ & $-8$ &
 $8$\\
 \hline
\end{tabular}
\end{center}.\\

Hence, plugging the results shown in the table for the coefficients
into \eq{BSL},  the result of taking the sum over the amplitudes for
the sub-processes shown in \fig{BLG} gives a gauge invariant
expression, and in the limit that $d\rightarrow\,4$, the
divergencies cancel exactly, such that the expression of \eq{BSL}
reduces to

\bea
A_b^{\mn}\,&=&\,A_b\,\Lb\,q_1^{\nu}q_2^{\mu}-\,\frac{M_H^2}{2}g^{\mn}\,\Rb\,\,\,\,\,\,\,\,\,\mbox{where}\,\,\,\,\,\,A_b\,=\,4\,\frac{\alpha_{em}2^{\frac{1}{4}}G_F^{\frac{1}{2}}}{\pi}\,\label{BSL2}
\eea

 \FIGURE[h]{ \centerline{ \epsfig{file=
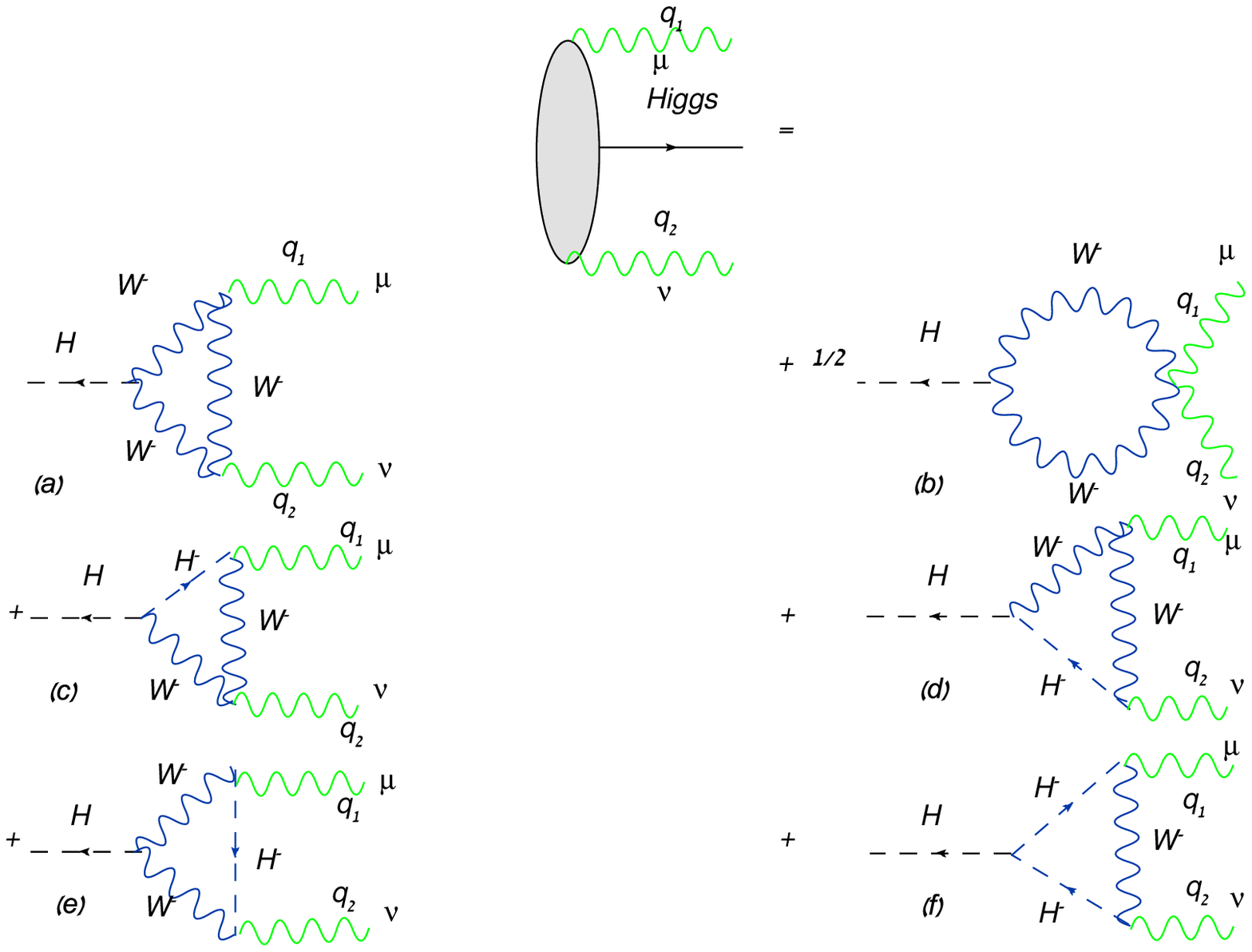,width=160mm,height=105mm}} \label{BL} }

\FIGURE[h]{ \centerline{ \epsfig{file=
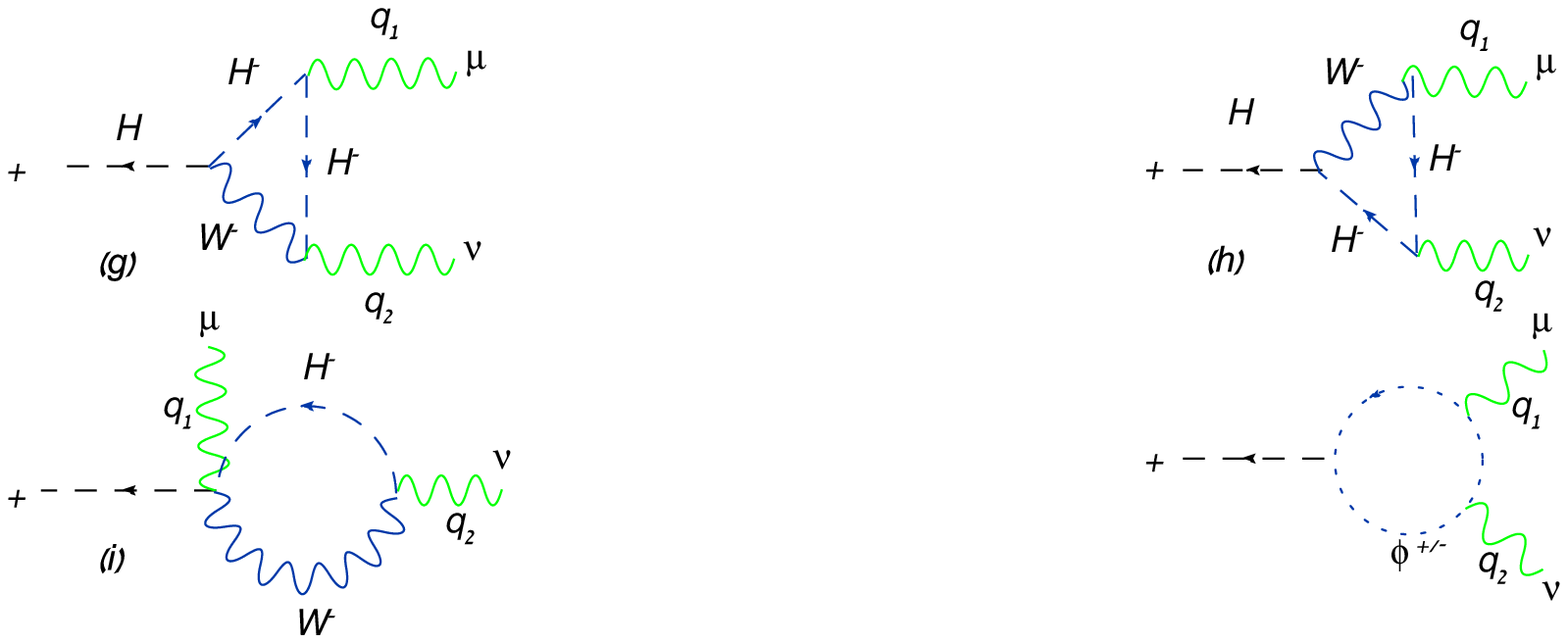,width=160mm,height=50mm}} \caption{Boson loop subprocess for
Higgs production in $\gamma\gamma$ fusion} \label{BLG} }
\newpage

It should be noted from the results for the amplitude of the
subprocess of \fig{BLG} (a), the subprocess
$\gamma\gamma\,\rightarrow\,\mbox{W triangle}\,\rightarrow\,H$,
interferes destructively with the subprocess
$\gamma\gamma\,\rightarrow\,\mbox{fermion triangle}\,\rightarrow\,H$
shown in \fig{leptr}.

\subsection{The cross section for central exclusive Higgs production
through $\gamma\gamma$ fusion}

Now that the amplitudes for the sub-processes
$\gamma\gamma\,\rightarrow\,\mbox{fermion triangle}\,\rightarrow\,H$
and $\gamma\gamma\rightarrow\,\mbox{boson loop}\,\rightarrow\,H$
have been calculated, the results can be plugged into \eq{E:35aa} to
derive the cross section, for central exclusive Higgs production
through $\gamma\gamma$ fusion. The result, taking into account all
possible sub-processes shown in \fig{leptr} and \fig{BLG} is found
to be

 \beq
\sigma^{\mbox{exc}}_{\gamma\gamma}\Lb\,\mbox{p+p}\rightarrow\,\mbox{p+H+p}\Rb\,=\,0.1\,\mbox{fb}
\eeq

\section{Conclusion}
\label{sec:con}

The results of this paper are summarized in the table below.
$\sigma^{\mbox{exc}}$ is the exclusive cross section, which includes
multiplication by a factor for the survival probability, for central
exclusive Higgs production. The results are given for the mechanisms
$pp\,\rightarrow\,\gamma\gamma\,\rightarrow\,H$ and
$pp\,\rightarrow\,\p\p\,\rightarrow\,H$. Note that in these results,
the cross section for central exclusive Higgs production in the case
of $\gamma\gamma$ fusion, is multiplied by a factor for the survival
probability of $1$. This is because in the case of photon exchange,
there is no hard re-scattering to suppress, and the large rapidity
gaps between the Higgs and the two emerging protons are
automatically present.
\\

\begin{center}\begin{tabular}{|c|c|r|}
    \hline
process &  $<\,\vert\,S^2\,\vert\,>$    &   $\sigma^{\mbox{exc}}\,\mbox{(fb)}$ \\
    \hline
$\p\p\,$   & 0.023 & 2.7  \\
$\p\p\,$  & 0.004 & 0.47 \\
$\gamma\gamma$  & 1 & 0.1\\ \hline
\end{tabular}
\end{center}.\\

The results show that, taking the survival probability to be $0.02$,
which is the value used in ref. \cite{kmr}, then the result for
$\sigma^{\mbox{exc}}_{\p\p}$ for central exclusive Higgs production
at the LHC, almost agrees with the prediction of ref. \cite{kmr},
(which was $3\,\mbox{fb}$). However, if the survival probability is
an order of magnitude smaller as predicted in ref. \cite{1}, then
$\sigma^{\mbox{exc}}_{\p\p}$ will be an order of magnitude smaller
and, it becomes competitive with
$\sigma^{\mbox{exc}}_{\gamma\gamma}$ for central exclusive Higgs
production at the LHC.

\section{Acknowledgements}
\label{sec:ack}

This paper is dedicated to the memory of Grandpa Herman, the
Pindenjara April 15th 1924 - January 2nd 2007.  I would like to
thank E. Levin for helpful advice in writing this paper. I would
also like to thank E. Gotsman, A. Kormilitzin, A. Prygarin for
fruitful discussions on the subject. This research was supported in
part  by the Israel Science Foundation, founded by the Israeli
Academy of Science and Humanities, by a grant from the Israeli
ministry of science, culture \& sport  \& the Russian Foundation for
Basic research of the Russian Federation,   and by the BSF grant \
20004019.

\renewcommand{\thesection}{A-\arabic{section}}
\setcounter{section}{0}

\renewcommand{\thesubsection}{A-\arabic{subsection}}
\setcounter{subsection}{0}
\renewcommand{\theequation}{A-1-\arabic{equation}}
\setcounter{equation}{0}  

\renewcommand{\thesection}{A-\arabic{section}}
\setcounter{section}{0}
\appendix
\section{Appendix}
 \label{sec:app}
\subsection{Evaluation of the integral over the anomalous dimensions
$\gamma_{1}$ and $\gamma_{2}$ of the momentum $Q^2$ in the gluon
density function} \label{sec:M}

The Born amplitude was calculated in \eq{E:2.4}, in terms of the
gluon density as a function of the anomalous dimensions $\gamma_{1}$
and $\gamma_{2}$, for the two gluon ladders in \fig{dd}. One now
needs to integrate over $\gamma_1$ and $\gamma_2$, and also over the
momentum in the t-channel gluon, namely $Q$. Altogether the
necessary integrations take the form
\begin{align}
M_{\p\p}\Lb\,\mbox{p+p}\rightarrow\,\mbox{p+H+p}\Rb\,=A\pi{}^{3}s\int{}d\gamma_{1}d\gamma_{2}\int\,\frac{dQ_{\bot}^{2}}{Q_{\bot}^{4}}e^{-S\left(k_{\bot}^{2},E_{\bot}^{2}\right)}f\left(x_{1},Q^{2}\right)f\left(x_{2},Q^{2}\right)
\label{E:A21}
\end{align}

Firstly, the integral over $Q_{\bot}$ is evaluated using the
steepest descent technique. The gluon density is given by
\begin{equation}
f\left(Q^{2},x_{1,2}\right)=2\left(Q^{2}\right)^{\gamma{}_{1,2}}e^{\omega\left(\gamma_{1,2}\right)\ln\frac{s_{1,2}}{s_{0}}}\label{E:B1}
\end{equation}

where
$\ln\Lb\,\frac{1}{x_{1,2}}\Rb\,\sim\ln\,\Lb\,\frac{\,s_{1,2}}{s_{0}}\Rb\,$,
where $s_{0}\sim1GeV$. This comes from the BFKL ladder gluon
exchange (see \fig{dd}), while the coefficient was taken from MRST -
NLO - 2002 data (see Ref/\cite{MRST}).
$\omega\left(\gamma_{1,2}\right)$ is the BFKL kernel defined as

\beq
\omega\Lb\gamma_{1,2}\Rb\,=\bas\chi\Lb\gamma_{1,2}\Rb\,=\bas\Lb\psi\Lb\,1\Rb\,-\psi\Lb\,\gamma_{1,2}\Rb\,-\psi\Lb\,1-\gamma_{1,2}\Rb\,\Rb\,\label{E:B2}
\eeq

where $\psi\Lb\,f\Rb$ is the digamma function and
$\psi\Lb\,f\Rb\,=\frac{d\Gamma\Lb\,f\Rb\,}{df}$. In \eq{E:B1},
$S\Lb\,k_{\bot}^2,E_{\bot}^2\Rb\,$ is the Sudakov form factor with
the typical value \cite{4,6}
$S\left(Q_{\bot}^{2},E_{\bot}\right)=\frac{3\alpha_{s}}{4\pi{}}\left(\ln{}\left(\frac{E_{\bot}^{2}}{Q_{\bot}^{2}}\right)\right)^{2}$,
in the notation that $E_{\bot}=\frac{M_{H}}{2}$. Using this
substitution \eq{E:A21} then becomes

\bea
&M_{\p\p}\Lb\,\mbox{p+p}\rightarrow\,\mbox{p+H+p}\Rb\,&=4A\pi{}^{3}s\int^{\infty}_{-\infty}d\gamma_{1}d\gamma_{2}\int{}\frac{dQ_{\bot}^{2}}{Q_{\bot}^{2}}\nonumber\\
&
&\times\exp\Lb-\phi{}\left(Q_{\bot}^{2}\Rb\Rb\exp\Lb\omega\Lb\,\gamma_{1}\Rb\,\ln\,\frac{s_{1}}{s_0}+\omega\Lb\,\gamma_{2}\Rb\,\ln\,\frac{s_{2}}{s_0}\Rb
\label{E:A22}\\
\mbox{where}&\phi\left(Q_{\bot}^{2}\right)&=\frac{3\alpha_{s}}{4\pi{}}\left(\ln{}\left(\frac{E_{\bot}^{2}}{Q_T^2}\right)\right)^{2}-\left(\gamma{}_{1}+\gamma{}_{2}-1\right)\ln{}Q_{\bot}^{2}\label{E:A23}
\eea

Differentiating the right hand side of \eq{E:A23} with respect to
$\ln\,Q_{\bot}^{2}$, one sees that $\phi$ has a saddle point at
$\ln\,Q_{\bot}^2=\ln{}\frac{M_{H}^{2}}{4}+\frac{2\pi{}}{3\alpha_{s}}\left(\gamma_{1}+\gamma_{2}-1\right)$.
Hence, changing the integration variable to $u=\ln\,Q_{\bot}^2$, and
expanding $\phi$ around the saddle point, \eq{E:A22} can be written
as

\bea
&M_{\p\p}\Lb\,\mbox{p+p}\rightarrow\,\mbox{p+H+p}\Rb\,&=4A\pi^{3}se^{-\phi\left(u_{0}\right)}\,\int{}d\gamma_{1}d\gamma_{2}
\int{}du\,e^{\Lb-\frac{1}{2}\left(u-u_{0}\right)^{2}\frac{d^2\phi\left(u_{0}\right)}{du^{2}}\Rb\,}
e^{\Lb\omega\Lb\,\gamma_{1}\Rb\,\ln\,\frac{s_{1}}{s_0}+\omega\Lb\,\gamma_{2}\Rb\,\ln\,\frac{s_{2}}{s_0}\Rb\,}\label{E:A30}\\
\mbox{where}&u_0&=\ln{}\frac{M_{H}^{2}}{4}+\frac{2\pi{}}{3\alpha_{s}}\left(\gamma_{1}+\gamma_{2}-1\right)\label{u0}\eea

Now the right hand side of \eq{E:A30} has reduced to a Gaussian
integral over $u$, which can be evaluated by the steepest descent
technique, to give the expression

\begin{align}
M_{\p\p}\Lb\,\mbox{p+p}\rightarrow\,\mbox{p+H+p}\Rb\,=&4A\pi^{4}s\Lb\frac{4}{3\alpha_s}\Rb^{\frac{1}{2}}\,\int{}d\gamma_{1}d\gamma_{2}\exp\Lb\left(\gamma_{1}+\gamma_{2}-1\right)\left(\frac{\pi{}}{3\alpha_{s}}\left(\gamma_{1}+\gamma_{2}-1\right)+\ln{}\frac{M_{H}^{2}}{4}\right)\Rb\notag\\
&\times\exp\Lb\,\omega\Lb\,\gamma_{1}\Rb\,\ln\,\frac{s_{1}}{s_0}+\omega\Lb\,\gamma_{2}\Rb\,\ln\,\frac{s_{2}}{s_0}\Rb\label{E:A31}
\end{align}

The BFKL function $\omega\Lb\,\gamma\Rb\,$ has a saddle point at
$\gamma=\frac{1}{2}$. Near to this point $\omega\Lb\,\gamma\Rb\,$
can be written as

\beq
\omega\Lb\gamma_{1,2}\Rb=\omega\Lb\frac{1}{2}\Rb+\frac{1}{2}\Lb\gamma_{1,2}-\frac{1}{2}\Rb^2\omega\,"\Lb\frac{1}{2}\Rb\label{E:B3}\eeq\,

Hence using \eq{E:B3}, \eq{E:A31} can be reduced to

\begin{align}
M_{\p\p}\Lb\,\mbox{p+p}\rightarrow\,\mbox{p+H+p}\Rb\,=&4A\pi^{4}s\Lb\frac{4}{3\alpha_s}\Rb^{\frac{1}{2}}\,\int{}d\gamma_{1}d\gamma_{2}\exp\Lb\,f\Lb\,\gamma_1,\gamma_2\Rb\,\Rb\label{E:B4}\end{align}

where the function $f\Lb\,\gamma_1,\gamma_2\Rb\,$ has the form

\begin{align}
f\Lb\gamma_{1},\gamma_{2}\Rb=&\omega\Lb\,\frac{1}{2}\Rb\,\ln\,\frac{s_{1}s_{2}}{s_0^2}+\frac{1}{2}\Lb\gamma_{1}-\frac{1}{2}\Rb^2\omega\,"\,\Lb\,\frac{1}{2}\Rb\ln\,\frac{s_{1}}{s_0}+\frac{1}{2}\Lb\gamma_{2}-\frac{1}{2}\Rb^2\omega\,"\,\Lb\,\frac{1}{2}\Rb\,\ln\,\frac{s_{2}}{s_0}\notag\\
&+\left(\gamma_{1}+\gamma_{2}-1\right)\left(\frac{\pi{}}{3\alpha_{s}}\left(\gamma_{1}+\gamma_{2}-1\right)+\ln{}\frac{M_{H}^{2}}{4}\right)\label{E:B5}
\end{align}

This function has a saddle point with respect to $\gamma_{1}$ given
by

\beq \gamma^{sp}_{1}
=\frac{-\frac{2\pi}{3\alpha_{s}}\Lb\gamma_{2}-1\Rb-\ln\frac{M_{H}^{2}}{4}+\frac{1}{2}\omega\,"\Lb\frac{1}{2}\Rb\ln\,\frac{s_{1}}{s_0}}{\Lb\frac{2\pi}{3\alpha_{s}}+\omega\,"\Lb\frac{1}{2}\Rb\ln\,\frac{s_{1}}{s_0}\Rb}\label{E:B6}
\eeq

Hence, expanding  $f\Lb\,\gamma_1,\gamma_2\Rb\,$ around
$\gamma_1^{sp}$, the integration over $\gamma_1$ is evaluated using
the steepest descent technique to give the expression

\begin{align}
M_{\p\p}\Lb\,\mbox{p+p}\rightarrow\,\mbox{p+H+p}\Rb\,=&2A\pi^{4}s\Lb\frac{4\pi}{3\alpha_s}\Rb^{\frac{1}{2}}\,\int{}d\gamma_{2}\frac{\exp\Lb\,f\Lb\gamma_{1}^{sp},\gamma_{2}\Rb\Rb}{\sqrt{-\frac{1}{2}\Lb\frac{2\pi}{3\alpha_{s}}+\ln\,\frac{M_{H}^{2}}{4}+\omega\,"\Lb\frac{1}{2}\Rb\ln\,\frac{s_{1}}{s_0}\Rb}}\label{E:B8}
\end{align}

Now the function $f\Lb\gamma_{1}^{sp},\gamma_{2}\Rb$  has a saddle
point with respect to $\gamma_{2}$ given by

\begin{align}
\gamma_{2}^{sp}=&\frac{1}{2}-\frac{\ln\frac{M_{H}^{2}}{4}\omega\,"\Lb\frac{1}{2}\Rb\ln\frac{s_{1}}{s_0}}{\frac{2\pi}{3\alpha_s}\omega\,"\Lb\frac{1}{2}\Rb\ln\,\frac{s_1s_2}{s_0^2}+\omega\,"\Lb\frac{1}{2}\Rb\ln\,\frac{s_{1}s_{2}}{s_0^2}}\sim\frac{1}{2}-\frac{\ln\,\frac{M_{H}^{2}}{4}}{\Lb\,1+\frac{4\pi}{3\alpha_{s}}\Rb\,\omega\,"\Lb\frac{1}{2}\Rb\,\ln\,\frac{s_{1}}{s_0}}\label{E:B8a}
\end{align}

Here in the second line it is assumed that $s_1\sim\,s_2$. For large
$s_1$ and $s_2$,  $\gamma_2^{sp}$ is approximately $\frac{1}{2}$.
Using the same method as above, expanding
$f\Lb\,\gamma_1^{sp},\gamma_2\Rb\,$ around $\gamma_{2}^{sp}$, the
integration over $\gamma_2$ is evaluated using the steepest descent
technique for \eq{E:B8}, to give the result

\bea
&M_{\p\p}\Lb\,\mbox{p+p}\rightarrow\,\mbox{p+H+p}\Rb\,&=A\pi^{4}s\Lb\frac{4\pi^2}{3\alpha_s}\Rb^{\frac{1}{2}}\,\frac{-2\exp\Lb\,f\Lb\gamma_{1}^{sp},\gamma^{sp}_{2}\sim\frac{1}{2}\Rb\Rb}{\Lb\frac{2\pi}{3\alpha_{s}}+\ln\,\frac{M_{H}^{2}}{4}+\omega\,"\Lb\frac{1}{2}\Rb\ln\,\frac{s_{1}}{s_0}\Rb}\label{E:B10}\\
\mbox{where}&f\Lb\,\gamma_{1}^{sp},\gamma_{2}^{sp}\Rb\,&=-\frac{\Lb\ln\,\frac{M_{H}^{2}}{4}\Rb^2}{\omega\,"\Lb\frac{1}{2}\Rb\ln\,\frac{s_{1}}{s_0}}\Lb\frac{1}{2}-\frac{\frac{\pi}{3\alpha_{s}}}{\omega\,"\Lb\frac{1}{2}\Rb\ln\,\frac{s_{1}}{s_0}}\Rb\label{E:B11}\eea

\newpage

\subsection{Feynman rules for the standard electro-weak theory}
\FIGURE[h]{ \centerline{\epsfig{file=
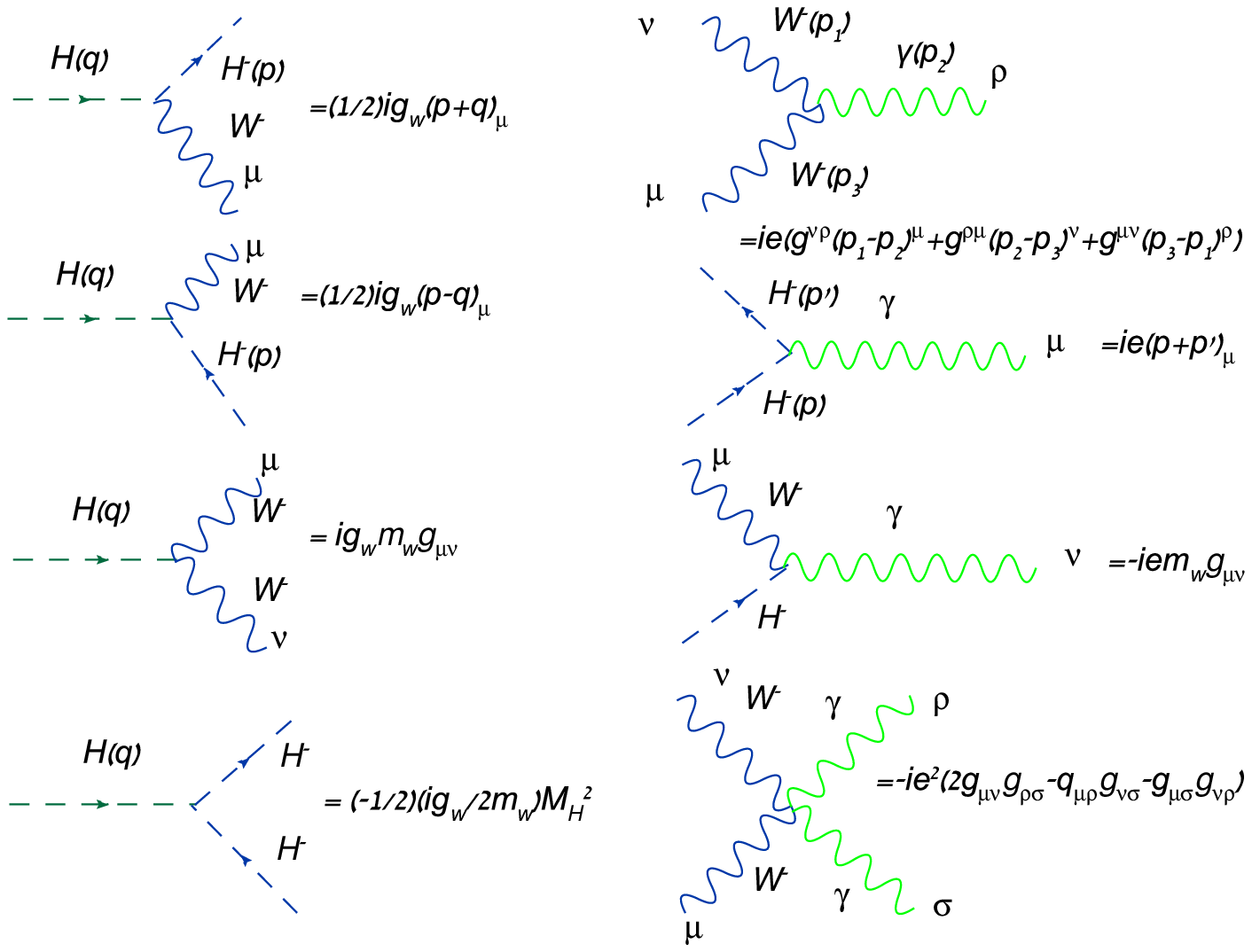,width=150mm,height=115mm}} \caption{} \label{feyn}}

\FIGURE[h]{ \centerline{\epsfig{file=
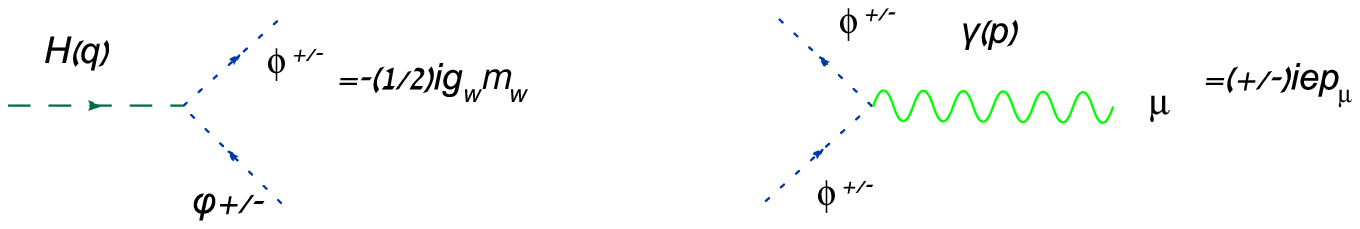,width=140mm,height=30mm}} \caption{Feynman rules in the
standard electroweak theory} \label{feyng}}

\renewcommand{\theequation}{A-3-\arabic{equation}}

\subsection{Evaluation of the integral over the momentum in the
fermion triangle loop} \label{sec:Elq} \setcounter{equation}{0}

The amplitude for the fermion triangle subprocess, summed over all
quark flavours $q\,=\,\Lb\,u\,,\,d\,,\,s\,,\,c\,,\,t\,,\,b\,\Rb\,$
and lepton flavours $L\,\,=\Lb\,e\,,\,\mu\,,\,\tau\,\Rb\,$, for the
mechanism $\gamma\gamma\,\mbox{fermion triangle}\,\rightarrow\,H$
was found in equation \eq{cf2} to take the form

\bea A^{\mu\nu}_f
&=&2\sum_f4\pi{}\alpha{}_{em}\left(G_{F}\sqrt{2}\right)^{\frac{1}{2}}\int{}\frac{d^{d}\tilde{l}}{\left(2\pi{}\right)^{d}}\int^{1}_{0}dx\int^{1-x}_{0}dy\frac{8m_{f}^2\Lb\,\Lb\,q_1^{\mu}q_2^{\nu}-\frac{M_H^2}{2}g^{\mn}\Rb\,\Lb\,1-2x\,y\Rb\,-2x\,y\,\frac{M_H^2}{2}g^{\mn}+m_f^2g^{\mn}\Rb\,}{\left(\tilde{l}^{2}-\Delta_f\Lb\,x,y\Rb\,\right)^{3}}\notag\\
&+&2\sum_f4\pi{}\alpha{}_{em}\left(G_{F}\sqrt{2}\right)^{\frac{1}{2}}\int{}\frac{d^{d}\tilde{l}}{\left(2\pi{}\right)^{d}}\int^{1}_{0}dx\int^{1-x}_{0}dy\frac{8m_{f}^2\Lb\,4\tilde{l}^{\mu}\tilde{l}^{\nu}-\tilde{l}^2\,g^{\mn}\Rb\,}{\left(\tilde{l}^{2}-\Delta_f\Lb\,x,y\Rb\,\right)^{3}}\,\,\,\,\,\,\mbox{where}\notag\\
\,\Delta_f\,&=&\,m_f^2-\,M_H^2\,xy\,\,\,\,\,\,\,\,\,\,\mbox{and}\,\,\,\,\,\,\sum_{f}=\,\sum_{q}\,+\sum_{L\,=\,e\,,\,\mu\,,\,\tau}\label{cf2a}\eea

where $\sum_q$ denotes the sum over all six quark flavours
$q\,=\,u\,,\,d\,,\,s\,,\,c\,,\,t\,,\,b$ and
$\sum_{L\,=\,e\,,\,\mu\,,\,\tau}$ denotes the sum over all three
lepton flavours $L\,=\,e\,,\,\mu\,,\,\tau$.

From the numerator of the integrand, on the RHS of \eq{cf2a}, one
can see that there is a gauge invariant term
$\Lb\,q_1^{\mu}q_2^{\nu}-\frac{M_H^2}{2}g^{\mn}\Rb\,\Lb\,1-4x\,y\Rb\,$,
and the numerator in the integrand of the second line on the RHS
gives a vanishing contribution to the integration over $\tilde{l}$,
for $d\rightarrow\,4$. However one is still left with the terms
$-2xy\frac{M_H^2}{2}g^{\mn}+m_f^2g^{\mn}$ in the numerator of the
integrand, which are certainly not gauge invariant. However,
conveniently this non gauge invariant piece is exactly equal to
$\Delta_f\,\Lb\,x,y\Rb\,$ introduced in \eq{cf}. To deal with this
non gauge invariant piece, it is useful to use dimensional
regularisation, when integrating over the $\tilde{l}^2$ term in the
numerator of the integrand on the second line. In this approach, one
initially integrates over $d+\epsilon{}$ dimensions in the limit
that $\epsilon{}\rightarrow{}0$ and $d\rightarrow\,4$. In this way
the non gauge invariant terms disappear. Hence, evaluating the
integral over $\tilde{l}$ on the RHS of \eq{cf2a} gives

\bea A^{\mu\nu}_f
&=&2\sum_f4\pi{}\alpha{}_{em}G_{F}^{\frac{1}{2}}2^{\frac{1}{4}}\frac{\Lb\,-1\Rb^3\Gamma\Lb\,3\,-\frac{d}{2}\Rb\,}{\Gamma\Lb\,3\Rb\,\Lb\,4\pi\Rb^{\frac{d}{2}}}\int^{1}_{0}dx\int^{1-x}_{0}dy\frac{8m_{f}^2\Lb\,\Lb\,q_1^{\mu}q_2^{\nu}-\frac{M_H^2}{2}g^{\mn}\Rb\,\Lb\,1-4x\,y\Rb\,+\Delta_f\Lb\,x,y\Rb\,g^{\mn}\Rb\,}{\Delta_f\Lb\,x,y\Rb\,}\notag\\
&+&2\,\lim_{\epsilon\,\to\,0}\sum_f4\pi{}\alpha{}_{em}G_{F}^{\frac{1}{2}}2^{\frac{1}{4}}\frac{\Lb\,-1\Rb^{3+1}\Gamma\Lb\,3\,-\frac{d+\epsilon}{2}-1\Rb\,}{2\Gamma\Lb\,3\Rb\,\Lb\,4\pi\Rb^{\frac{d+\epsilon}{2}}}\int^{1}_{0}dx\int^{1-x}_{0}dy\,8m_{f}^2\Lb\,4g^{\mn}-\Lb\,d+\epsilon\Rb\,\,g^{\mn}\Rb\,\label{dr}
\eea

In the limit that $\epsilon\,\rightarrow\,0$ and $d\rightarrow\,4$,
the gamma function
$\Gamma\Lb\,3-\frac{d+\epsilon}{2}-1\Rb\,\rightarrow\,\frac{2}{\epsilon}$
and the RHS of \eq{dr} reduces to

\bea A^{\mu\nu}_f
&=&\,-2\sum_f4\pi{}\alpha{}_{em}G_{F}^{\frac{1}{2}}2^{\frac{1}{4}}\frac{1}{2\,\Lb\,4\pi\Rb^{2}}\int^{1}_{0}dx\int^{1-x}_{0}dy\frac{8m_{f}^2\Lb\,\Lb\,q_1^{\mu}q_2^{\nu}-\frac{M_H^2}{2}g^{\mn}\Rb\,\Lb\,1-4x\,y\Rb\,+\Delta_f\Lb\,x,y\Rb\,g^{\mn}\Rb\,}{\Delta_f\Lb\,x,y\Rb\,}\notag\\
&+&2\,\sum_f4\pi{}\alpha{}_{em}G_{F}^{\frac{1}{2}}2^{\frac{1}{4}}\frac{1}{\epsilon}\frac{1}{2\Lb\,4\pi\Rb^{2}}\int^{1}_{0}dx\int^{1-x}_{0}dy\,8m_{f}^2\,\epsilon\,g^{\mn}\label{dr2}
\eea

Thus, after canceling $\epsilon$ in the numerator and the
denominator in the second line on the RHS of \eq{dr2}, the second
line exactly cancels the non gauge invariant part of the integrand
on the first line. Hence, one is left with the purely gauge
invariant expression

\bea A^{\mu\nu}_q
&=&\,-2\,\frac{\alpha{}_{em}G_{F}^{\frac{1}{2}}2^{\frac{1}{4}}}{\pi}\,\Lb\,q_1^{\mu}q_2^{\nu}-\frac{M_H^2}{2}g^{\mn}\Rb\,\sum_f\,I_f\label{dr3}
\eea

where $I_f$ is the only remaining integral to evaluate, which takes
the form \beq
\sum_f\,I_f\,=\,\sum_{f\,=\,\mbox{$u\,,\,d\,,\,s\,,\,c\,,\,t\,,\,b\,$}\,,\mbox{$e\,,\,\mu\,,\,\tau\,$}}m_f^2\int^1_0dx\int^{1-x}_0\,dy\frac{1-4xy}{\Delta_f\Lb\,x,y\Rb}\,\,\,\,\,\,\,\,\,\mbox{where}\,\,\,\,\,\Delta_f\Lb\,x,y\Rb\,=m_f^2-M_H^2xy
\label{int}\eeq

To evaluate this integral, this are two cases to consider, namely
(1) when $m_f^2\,\gg\,M_H^2$, which is true for the top quark when
$m_f\,=\,m_{t}\,=175\,\mbox{GeV}$, and (2) when $m_f^2\,\ll\,M_H^2$,
which is true for all the rest of the fermions listed in the table.
Therefore $\sum_f\,I_f$ can be separated into two parts, namely

\bea \sum_f\,I_f\,=\,I_{t}\,+\,\sum_{f\,\neq\,t}\,I_f\label{neq}
\eea

For case (1), where $m_f\,=\,m_t\gg\,M_H\,$, one can write $I_t$ in
a more convenient way as

\bea
I_t\,&=&\frac{m_t^2}{M_H^2}\,\int^1_0dx\int^{1-x}_0\,dy\frac{1-4xy}{\frac{m_t^2}{M_H^2}-xy}\,=\,\frac{m_t^2}{M_H^2}\int^1_0dx\int^{1-x}_0dy\,\Lb\,4\,+\frac{1-4\frac{m_t^2}{M_H^2}}{\frac{m_t^2}{M_H^2}-xy}\Rb\,\notag\\
&=&\,\frac{m_t^2}{M_H^2}\Lb\,2\,-\,\Lb\,1-4\,\frac{m_t^2}{M_H^2}\Rb\,\int^1_0\frac{dx}{x}\ln\Lb\,1-\frac{M_H^2}{m_t^2}x\Lb\,1-x\Rb\,\Rb\,\Rb\,\notag\\
&=&\frac{m_t^2}{M_H^2}\Lb\,2+\,\Lb\,1-4\,\frac{m_t^2}{M_H^2}\Rb\,\int^1_0dx\Lb\,\frac{M_H^2}{m_t^2}\Lb\,1-x\Rb\,+\frac{1}{2}\Lb\,\frac{M_H^2}{m_t^2}\Rb^2\,x\Lb\,1-x\Rb^2\,+\frac{1}{3}\Lb\,\frac{M_H^2}{m_t^2}\Rb^3x^2\Lb\,1-x\Rb^3+.....\Rb\,\Rb\,\label{inte222}\notag\\
\eea

where in the last step, the logarithm was expanded in a Taylor
series around $x\,=\,0$. Evaluating the integral over $x$, and since
it is assumed that $M_H^2\ll\,m_t^2$, retaining  terms no smaller
than $\Lb\,\frac{M_H^2}{m_t^2}\Rb\,$, the RHS of \eq{inte222}
becomes

\bea
I_t\,&=&\frac{1}{3}\,\Lb\,1+\,\frac{7}{120}\frac{M_H^2}{m_t^2}+....\Rb\,\label{inte}\eea

For case (2), where $m_f\ll\,M_H$ which includes all the fermions in
the table except for the top quark, there are two possible regions
of integration, namely (I) when $M_H^2\,x\,y\,>\,m_f^2$ and (II)
when $M_H^2\,x\,y\,<\,m_f^2$. In the region where
$M_H^2\,x\,y>\,m_f^2\,$, the RHS of \eq{int} reduces to

\bea
\sum_{f\neq\,t}\,I_f^{\mbox{region (I) }}&=&-\,\sum_{f\neq\,t}\,\,\frac{m_f^2}{M_H^2}\int^1_{\frac{m_f^2}{M_H^2}}dx\int^{1-x}_{\frac{m_f^2}{M_H^2\,x}}\,dy\frac{1-4xy}{\,x\,y}\,\notag\\
&=&\,-\,\sum_{f\neq\,t}\,\frac{m_f^2}{M_H^2}\,\int^1_{\frac{m_f^2}{M_H^2}}\frac{dx}{x}\ln\,\Lb\,\frac{M_H^2}{m_f^2}x\Lb\,1-x\Rb\,\Rb\,+4\,\sum_{f\neq\,t}\,\frac{m_f^2}{M_H^2}\int^1_{\frac{m_f^2}{M_H^2}}\Lb\,1-x\,-\frac{m_f^2}{M_H^2}\frac{1}{x}\Rb\,\notag\\
&=&\,-\,\sum_{f\neq\,t}\,\frac{m_f^2}{2M_H^2}\ln^2\frac{M_H^2}{m_f^2}+\frac{m_f^2}{M_H^2}\mbox{polylog}\Lb\,2\,,x=1\Rb\,-\,\sum_{f\neq\,t}\,\frac{m_f^2}{M_H^2}\mbox{polylog}\Lb\,2\,,x=\,\frac{m_f^2}{M_H^2}\Rb\,\notag\\
&\,&+2\,\sum_{f\neq\,t}\,\Lb\,\frac{m_f^2}{M_H^2}+4\Lb\,\frac{m_f^2}{M_H^2}\Rb^2\ln\,\frac{m_f^2}{M_H^2}\Rb\,\label{intLep1}\eea

In the region (II) where $M_H^2\,x\,y<\,m_f^2\,$, the RHS of
\eq{int} reduces to

\bea
\sum_{f\neq\,t}\,I_f^{\mbox{region (II) }}\,&=&\,\sum_{f\neq\,t}\,\,m_f^2\,\int_{\frac{m_f^2}{M_H^2}}^{1}dx\int^{\frac{m_f^2}{M_H^2}\frac{1}{x}}_{0}\,dy\,\frac{\Lb\,1-4xy\Rb\,}{m_f^2}\,\,=\,\sum_{f\neq\,t}\,\int_{\frac{m_f^2}{M_H^2}}^{1}dx\,\frac{m_f^2}{M_H^2}\frac{1}{x}\Lb\,1-2\frac{m_f^2}{M_H^2}\Rb\,\notag\\
&=&\,\sum_{f\neq\,t}\,\,\frac{m_f^2}{M_H^2}\Lb\,1-2\frac{m_f^2}{M_H^2}\Rb\,\ln\frac{M_H^2}{m_f^2}\label{intLep2}
\eea

Hence, adding the contributions of \eq{intLep1} and \eq{intLep2} for
the contributions of region (I) and region (II) of the integral,
gives the result

\bea
\sum_{f\neq\,t}\,I_f\,&=&-\,\sum_{f\neq\,t}\,\frac{m_f^2}{M_H^2}\Lb\,1-6\,\frac{m_f^2}{M_H^2}-\frac{1}{2}\ln\frac{M_H^2}{m_f^2}\Rb\,\ln\frac{M_H^2}{m_f^2}\notag\\
&\,&+\,\sum_{f\neq\,t}\,\frac{m_f^2}{M_H^2}\Lb\,\mbox{polylog}\Lb\,2\,,x=1\Rb\,-\mbox{polylog}\Lb\,2\,,x=\frac{m_f^2}{M_H^2}\Rb\,+2\Rb\,
\label{intLep3}\\
&\approx\,&\,\sum_{f\neq\,t}\,\frac{1}{2}\frac{m_f^2}{M_H^2}\ln^2\frac{M_H^2}{m_f^2}\,\,\,\,\,\,\,\,\,\,\,\,\mbox{for}\,\,m_f\ll\,M_H\,\label{intLep3s}\eea

From \eq{intLep3s} and \eq{inte}, the result for the evaluation of
the integral $I_f$ has its main contribution from the top quark
triangle, such that

\beq
\sum_f\,I_f\,\approx\,I_t\,,\,=\,\frac{1}{3}\,\,\,\,\,\,\,\mbox{for}\,\,\,\,\,\,\,m_t\gg\,M_H\,\label{IR}
\eeq

Plugging this result into \eq{dr3} gives the final expression for
the amplitude of the fermion triangle subprocess shown in
\fig{leptr}, for the sum over all quark
$q\,=\,\Lb\,u\,,\,d\,,\,s\,,\,c\,,\,t\,,\,b\,\Rb\,$ contributions
and lepton contributions $L\,\,=\,e\,,\,\mu\,,\,\tau\,$ as

\bea A^{\mu\nu}_f
&=&\,A_f\,\Lb\,q_1^{\mu}q_2^{\nu}-\frac{M_H^2}{2}g^{\mn}\Rb\,\,\,\mbox{where}\,\,\,\,\,A_f\,=\,-\frac{2}{3}\frac{\alpha{}_{em}G_{F}^{\frac{1}{2}}2^{\frac{1}{4}}}{\pi}\,\label{dr3Lep22}
\eea

\def\ap#1#2#3{     {\it Ann. Phys. (NY) }{\bf #1} (19#2) #3}
\def\arnps#1#2#3{  {\it Ann. Rev. Nucl. Part. Sci. }{\bf #1} (19#2) #3}
\def\npb#1#2#3{    {\it Nucl. Phys. }{\bf B#1} (19#2) #3}
\def\plb#1#2#3{    {\it Phys. Lett. }{\bf B#1} (19#2) #3}
\def\prd#1#2#3{    {\it Phys. Rev. }{\bf D#1} (19#2) #3}
\def\prep#1#2#3{   {\it Phys. Rep. }{\bf #1} (19#2) #3}
\def\prl#1#2#3{    {\it Phys. Rev. Lett. }{\bf #1} (19#2) #3}
\def\ptp#1#2#3{    {\it Prog. Theor. Phys. }{\bf #1} (19#2) #3}
\def\rmp#1#2#3{    {\it Rev. Mod. Phys. }{\bf #1} (19#2) #3}
\def\zpc#1#2#3{    {\it Z. Phys. }{\bf C#1} (19#2) #3}
\def\mpla#1#2#3{   {\it Mod. Phys. Lett. }{\bf A#1} (19#2) #3}
\def\nc#1#2#3{     {\it Nuovo Cim. }{\bf #1} (19#2) #3}
\def\yf#1#2#3{     {\it Yad. Fiz. }{\bf #1} (19#2) #3}
\def\cpc#1#2#3{    {\it Comp. Phys. Commun. }{\bf #1} (19#2) #3}
\def\dis#1#2{      {\it Dissertation, }{\sf #1 } 19#2}
\def\dip#1#2#3{    {\it Diplomarbeit, }{\sf #1 #2} 19#3 }
\def\ib#1#2#3{     {\it ibid. }{\bf #1} (19#2) #3}
\def\jpg#1#2#3{        {\it J. Phys}. {\bf G#1}#2#3}
\bibliographystyle{amsplain}
\bibliography{}

\begin{thebibliography}{99}




\bibitem{1}
J.S. Miller, \emph{"Survivial probability for Higgs diffractive
production in high density QCD"} \emph{(in press)}\\
arxiv: hep-ph/0610427

\bibitem{kmr}
V.A. Khoze A.D.Martin M.G. Ryskin \emph{"Prospects for new physics
observations
in diffractive processes at the LHC and Tevatron"}\\
\emph{Eur. Phys. J.} {\bf C23} 311 - 327 (2002)\\
arxiv: hep-ph/0111078


\bibitem{4}
V. Khoze, A.Martin, M.Ryskin, \emph{"The rapidity gap Higgs signal
at the LHC"} \plb{401}{1997}{330-336}\\
arXiv:hep-ph/9701419


\bibitem{6}
V.Khoze, A.Martin, M.Ryskin, \emph{"Dijet hadroproduction with
rapidity gaps and QCD double logarithmic effects"}
\prd{56}{1997}{5867-5874}\\
arXiv:hep-ph/9705258

\bibitem{5}
G.Altarelli, G.Parisi, \npb{126}{1977}{298}


\bibitem{Rizz}
Thomas G. Rizzo, \emph{"Gluon final states in Higgs - Boson decay"},
\prd{22}{1980}{178}, Addendum-ibid \prd{22}{1980}{1824-1825}

\bibitem{2}
J. Ellis et al., \emph{"Higgs boson"} Nucl. Phys. \textbf{B106} 326-
331 (1976)



\bibitem{Higgs}
J. Ellis et al., \emph{"A phenomenological profile of the Higgs
boson"}\npb{106}{1976}{326-331}


\bibitem{daw}
S. Dawson, \emph{"Radiative corrections to Higgs boson production"}
\npb{359}{1991}{283-300}

\bibitem{22}
S. Bentvelsen, E. Laenen, P. Motylinski, \emph{"Higgs production
through gluon fusion at leading order"} NIKHEF 2005 - 007



















\end{thebibliography}
\label{sec:bib}

\end{document}